\documentclass[acmtog]{acmart}
\usepackage{booktabs} % For formal tables
\usepackage[percent]{overpic}
\usepackage{subfigure}

\usepackage{multirow}

\usepackage[ruled]{algorithm2e} % For algorithms

\SetAlFnt{\small}
\SetAlCapFnt{\small}
\SetAlCapNameFnt{\small}
\SetAlCapHSkip{0pt}

\acmJournal{TOG}
\newcommand{\f}[1]{\mathcal{#1}}
\newcommand{\p}{\partial}

\newcommand{\bs}[1]{\boldsymbol{#1}}

\newcommand{\dt}{\Delta t} 
\newcommand{\dx}{\Delta x}

\definecolor{teal}{RGB}{0,128,129}
\definecolor{ruby}{RGB}{224,17,95}
\definecolor{ffc}{RGB}{17,227,95}
\newcommand{\tx}[1]{\textcolor{teal}{#1}}

\begin{document}
% Title portion
\title{$A$-ULMPM: An Arbitrary Updated Lagrangian Material Point Method for Efficient Simulation of Solids and Fluids}
\author{Haozhe Su}\authornote{joint first authors}\affiliation{\institution{Rutgers University}}

\author{Tao Xue}\authornotemark[1]\affiliation{\institution{Rutgers University}}

\author{Chengguizi Han}\affiliation{\institution{Rutgers University}}

\author{Mridul Aanjaneya}\affiliation{\institution{Rutgers University}}

\renewcommand\shortauthors{Haozhe Su, Tao Xue, Chengguizi Han, and Mridul Aanjaneya}
% DO NOT ENTER AUTHOR INFORMATION FOR ANONYMOUS TECHNICAL PAPER SUBMISSIONS TO SIGGRAPH 2019!
%\author{Gang Zhou}
%\orcid{1234-5678-9012-3456}
%\affiliation{%
%  \institution{College of William and Mary}
%  \streetaddress{104 Jamestown Rd}
%  \city{Williamsburg}
%  \state{VA}
%  \postcode{23185}
%  \country{USA}}
%\email{gang_zhou@wm.edu}
%\author{Valerie B\'eranger}
%\affiliation{%
%  \institution{Inria Paris-Rocquencourt}
%  \city{Rocquencourt}
%  \country{France}
%}
%\email{beranger@inria.fr}
%\author{Aparna Patel}
%\affiliation{%
% \institution{Rajiv Gandhi University}
% \streetaddress{Rono-Hills}
% \city{Doimukh}
% \state{Arunachal Pradesh}
% \country{India}}
%\email{aprna_patel@rguhs.ac.in}
%\author{Huifen Chan}
%\affiliation{%
%  \institution{Tsinghua University}
%  \streetaddress{30 Shuangqing Rd}
%  \city{Haidian Qu}
%  \state{Beijing Shi}
%  \country{China}
%}
%\email{chan0345@tsinghua.edu.cn}
%\author{Ting Yan}
%\affiliation{%
%  \institution{Eaton Innovation Center}
%  \city{Prague}
%  \country{Czech Republic}}
%\email{yanting02@gmail.com}
%\author{Tian He}
%\affiliation{%
%  \institution{University of Virginia}
%  \department{School of Engineering}
%  \city{Charlottesville}
%  \state{VA}
%  \postcode{22903}
%  \country{USA}
%}
%\affiliation{%
%  \institution{University of Minnesota}
%  \country{USA}}
%\email{tinghe@uva.edu}
%\author{Chengdu Huang}
%\author{John A. Stankovic}
%\author{Tarek F. Abdelzaher}
%\affiliation{%
%  \institution{University of Virginia}
%  \department{School of Engineering}
%  \city{Charlottesville}
%  \state{VA}
%  \postcode{22903}
%  \country{USA}
%}

%\renewcommand\shortauthors{Zhou, G. et al}

\begin{abstract}
We present an arbitrary updated Lagrangian Material Point Method ($A$-ULMPM) to alleviate issues, such as the cell-crossing instability and numerical fracture, that plague state of the art Eulerian formulations of MPM, while still allowing for large deformations that arise in fluid simulations. Our proposed framework spans MPM discretizations from total Lagrangian formulations to Eulerian formulations. We design an easy-to-implement physics-based criterion that allows $A$-ULMPM to update the reference configuration \emph{adaptively} for measuring physical states including stress, strain, interpolation kernels and their derivatives. For better efficiency and conservation of angular momentum, we further integrate the APIC~\cite{jiang:2015:apic} and MLS-MPM~\cite{Hu:2018:Moving} formulations in $A$-ULMPM by augmenting the accuracy of velocity rasterization using both the local velocity and its first-order derivatives. Our theoretical derivations use a nodal discretized Lagrangian, instead of the weak form discretization in MLS-MPM~\cite{Hu:2018:Moving}, and naturally lead to a ``modified'' MLS-MPM in $A$-ULMPM, which can recover MLS-MPM using a completely Eulerian formulation.  
$A$-ULMPM does not require significant changes to traditional Eulerian formulations of MPM, and is computationally more efficient since it only updates interpolation kernels and their derivatives when large topology changes occur.
We present end-to-end 3D simulations of stretching and twisting hyperelastic solids, splashing liquids, and multi-material interactions with large deformations to demonstrate the efficacy of our novel $A$-ULMPM framework.
\end{abstract}

%
% The code below should be generated by the tool at
% http://dl.acm.org/ccs.cfm
% Please copy and paste the code instead of the example below.
%


%
% End generated code
%


\begin{teaserfigure}
    \vspace{-2mm}
    \includegraphics[width=\textwidth]{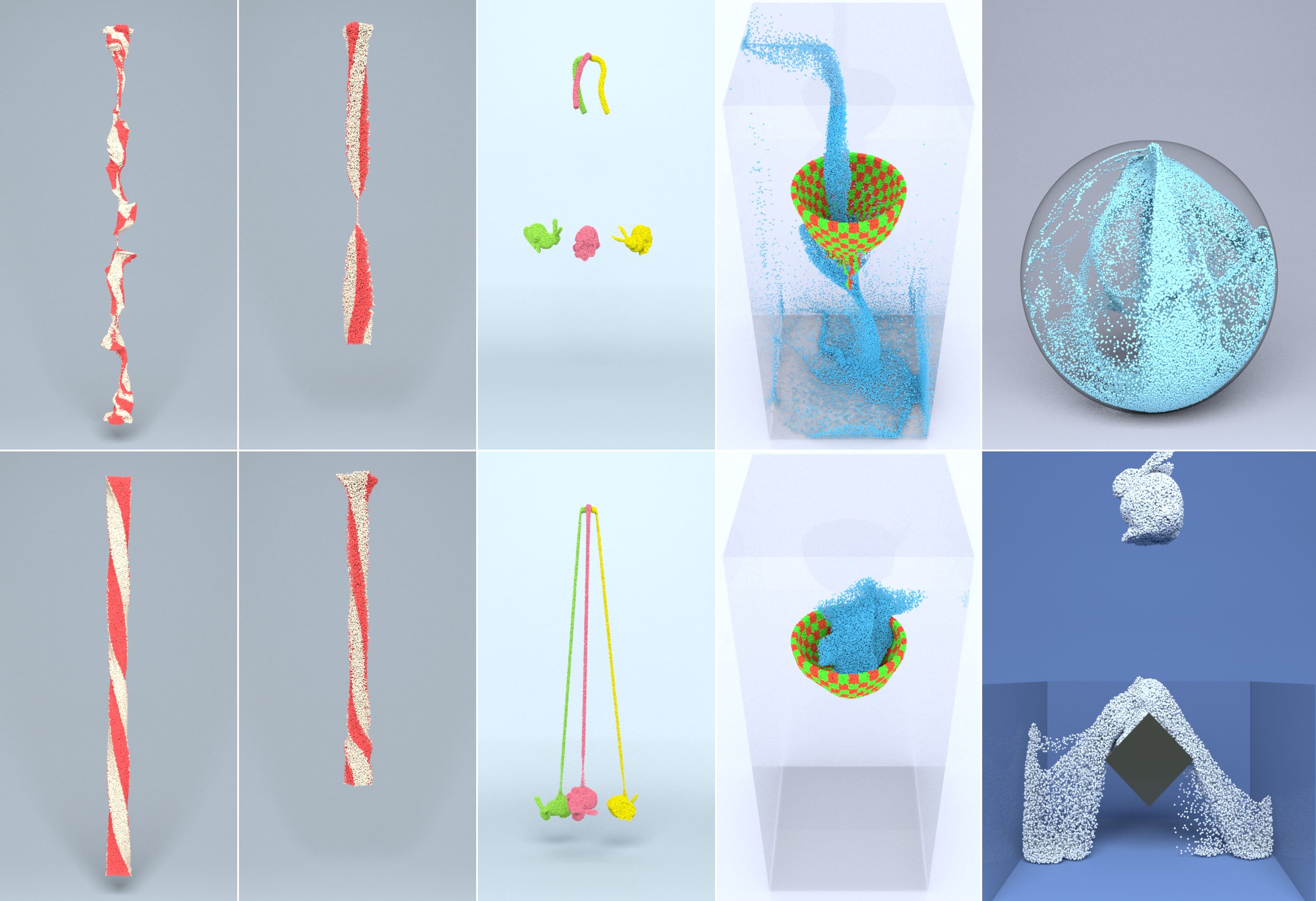}
    \vspace{-8mm}
    \caption{Our $A$-ULMPM framework (bottom) avoids issues such as numerical fracture and the cell-crossing instability that plague Eulerian approaches to MPM (top) and allows for large deformations in solid and fluid simulations while requiring $4.27\times$ -- $31.52\times$ less computational overhead for configuration updates.}
    \label{fig:teaser}
\end{teaserfigure}

\maketitle

\section{INTRODUCTION}
\label{sec:introduction}
The Material Point Method (MPM) family of discretizations~\cite{Sulsky:1994:Particle}, such as Fluid Implicit Particle (FLIP)~\cite{Brackbill:1988:Flip} and Particle-in-Cell (PIC)~\cite{Sulsky:1995:PIC}, emerged as an effective choice for simulating various materials and gained popularity in visual effects (VFX) for providing high-fidelity physics simulations of snow~\cite{Stomakhin:2013:MPMsnow}, sand~\cite{Klar:2016:Drucker,Daviet:2016:Semi-implicit}, phase change~\cite{Stomakhin:2014:Augmented,gao2018gpu}, viscoelasticity~\cite{Ram:2015:Material,yue:2015:mpm-plasticity,Su:2021:USOSVL}, viscoplasticity~\cite{Fang:2019:Silly}, elastoplasticity~\cite{Gao:2017:AGIMPM}, fluid structure interactions~\cite{Fang:2020:IQ-MPM}, fracture~\cite{Wolper:2019:CD-MPM,hegemann:2013:ductile}, fluid-sediment mixtures~\cite{Tampubolon:2017:Multi,gao2018animating}, baking and cooking~\cite{Ding:2019:Thermomechanical}, and diffusion-driven phenomena~\cite{XUE:2020:NF}.
In contrast to Lagrangian mesh-based methods, such as the Finite Element Method (FEM)~\cite{Zienkiewicz:1977:FEM,Sifakis:2012:FEM}, and pure particle-based methods, such as Smoothed Particle Hydrodynamics (SPH)~\cite{Desbrun:1996:SPH,Liu:2008:Overview}, MPM merges the advantages of both Lagrangian and Eulerian approaches and automatically supports dynamic topology changes such as material splitting and merging. 
It uses Lagrangian particles to carry material states, while the background grid acts as an Eulerian ``scratch pad'' for computing the divergence of stress and performing spatial/temporal numerical integration. 
The use of a background grid allows for regular numerical stencils, benefiting from cache-locality, while the use of particles avoids the numerical dissipation issues characteristics of Eulerian grid-based schemes.

\begin{figure*}[t]
\vspace*{-4mm}
\begin{center}
\includegraphics[width=\textwidth]{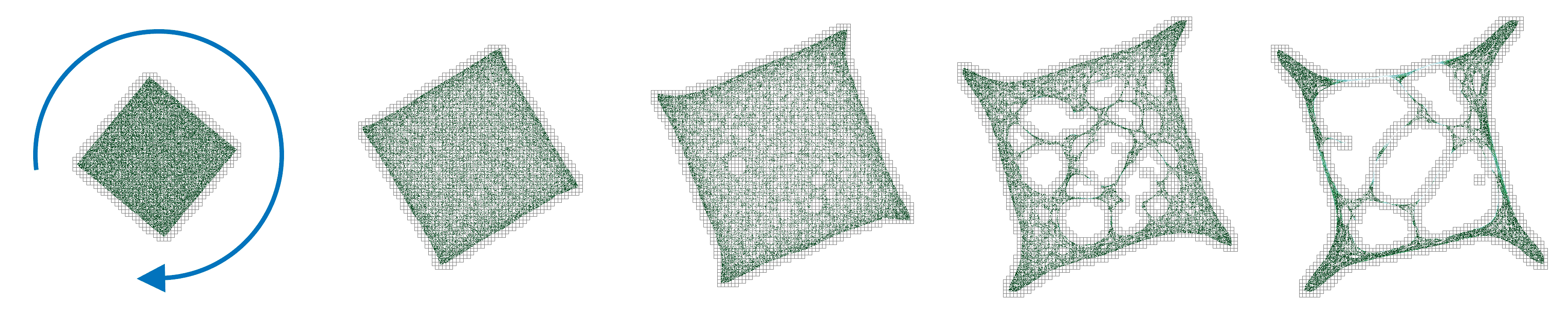}\\
\vspace{-1mm}
MLS-EMPM: $t=\{0.00833,\:0.01667,\:0.02500,\:0.03333,\:0.04167\}$s\\
\includegraphics[width=\textwidth]{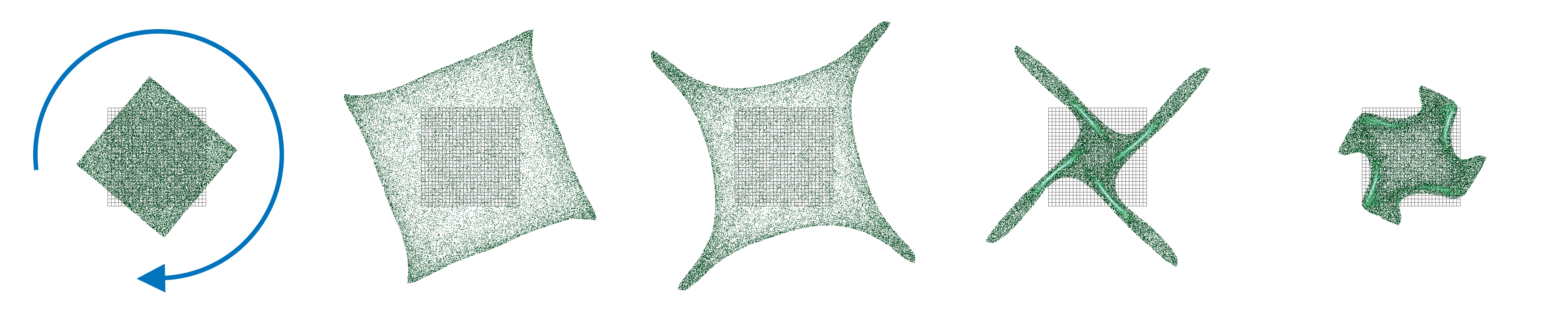}\\
\vspace{-1mm}
MLS-TLMPM: $t=\{0.00833,\:0.02500,\:0.04167,\:0.06250,\:0.07917\}$s\\
\includegraphics[width=\textwidth]{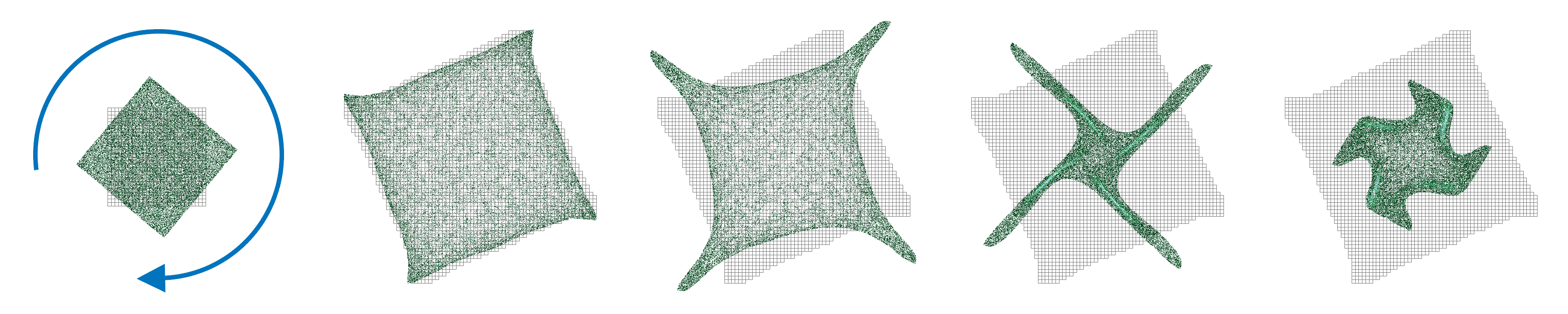}\\
\vspace{-3mm}
MLS-$A$-ULMPM: $t=\{0.00833,\:0.02500,\:0.04167,\:0.06250,\:0.07917\}$s\\
\vspace{-4mm}
\end{center}
\caption{\textbf{Eliminating numerical fracture.} Our $A$-ULMPM (bottom row) and TLMPM (middle row) for solid simulation capture the appealing hyperelastic rotation, preserve the angular momentum for long simulation periods, and completely eliminate numerical fracture.
Traditional EMPM (top row) suffers from severe numerical fracture in solid simulations when large deformations occur, while TLMPM (middle row) does not. 
 EMPM updates configurations simultaneously with respect to particles, while TLMPM does not update configurations at all. $A$-ULMPM only updates configurations whenever the shape changes reach to our predefined criterion (see equation~\eqref{eq:update_conf}). The background grid represents the configuration linking to the present particle dynamics.}
\vspace*{-4mm}
\label{fig:2D_rotating}
\end{figure*}

Conventional MPM discretization employs an \emph{Eulerian} formulation, which measures stress and strain and computes derivatives and integrals with respect to the Eulerian coordinates (i.e., the ``current'' configuration). 
Eulerian MPM (EMPM) has been acknowledged as a powerful tool for physics-based simulations~\cite{jiang2016material}, particularly if the system involves large deformations, such as fluid-like motion. 
However, it suffers from a number of shortcomings such as the \emph{cell-crossing instability}. 
Many studies have shown that when cell-crossing occurs, the MPM solutions can either be non-convergent or reduce the convergence rate when refining grid, with the spatial convergence rate varying between first and second order~\cite{W:2007:improved} (see Figure~\ref{fig:ConvPlot}).
The deficiency of cell-cross instability has been reduced in the latest MPM formulations, such as the Affine Particle-In-Cell (APIC) method~\cite{Jiang:2017:APIC}, the generalized interpolation MPM (GIMP)~\cite{Bardenhagen:2004:GIMPM, Gao:2017:AGIMPM}, and the Convected Particles Domain Interpolation (CPDI)~\cite{Sadeghirad:2011:CPDI}. 
Among them, the APIC approach has been widely adopted in computer graphics. The central idea behind APIC is to retain the filtering property of PIC, but reduce dissipation by interpolating more information, such as the velocity and its \emph{derivatives}, aiming to conserve linear and angular momentum. 
An improved APIC, namely PolyPIC, was proposed in~\cite{Fu:2017:PolyPIC} that allows for locally \emph{high-order approximations}, rather than approximations to the grid velocity field.
Later on, a \emph{moving least squares} MPM formulation (MLS-MPM)~\cite{Hu:2018:Moving} was developed by introducing the MLS technique to elevate the accuracy of the internal force evaluation and velocity derivatives. 
 
 Although EMPM discretizations have been improved to a certain degree via the aforementioned MLS techniques, they still suffer from \emph{numerical fracture} that occurs when particles move far enough from the cell where they are originally located, and a gap of one cell or more is created between them. 
 Such non-physical fracture depends only on the background grid resolution and is not related to any other issue that would ultimately limit the accuracy of the EMPM to model actual physical fracture of materials under large deformations (see Figure~\ref{fig:2D_rotating}), particularly in solid simulations.
 
\begin{figure*}[t]
\vspace*{-4mm}
\begin{center}
\includegraphics[width=\textwidth]{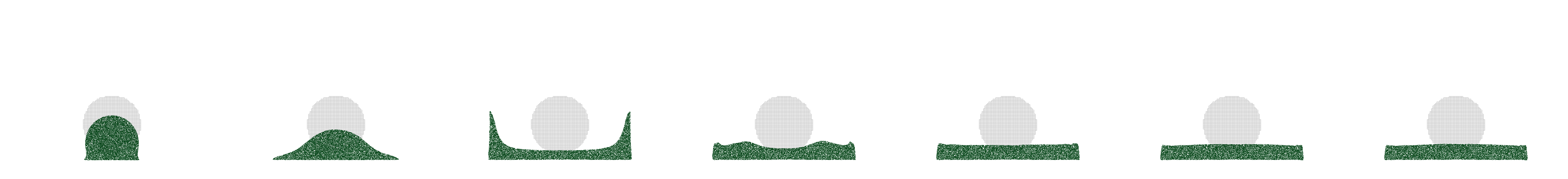}\\
\vspace{-1mm}
Kernel-TLMPM: $t=\{0.14167,\:0.20833,\:0.47083,\:0.74167,\:1.31667,\:1.88750,\:2.04583\}$s\\
\includegraphics[width=\textwidth]{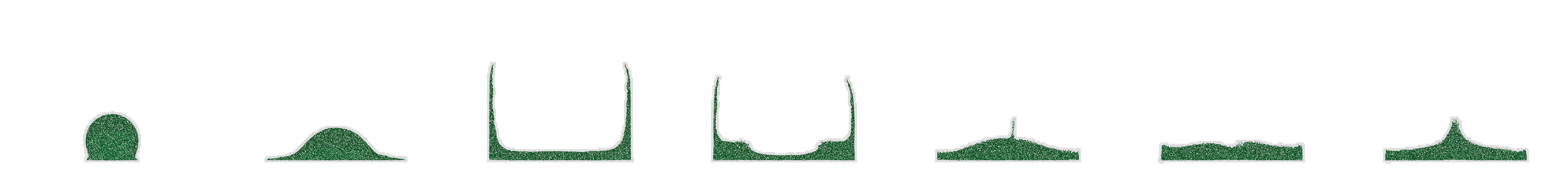}\\
\vspace{-1mm}
Kernel-EMPM: $t=\{0.14167,\:0.20833,\:0.47083,\:0.74167,\:1.31667,\:1.88750,\:2.06250\}$s\\
\includegraphics[width=\textwidth]{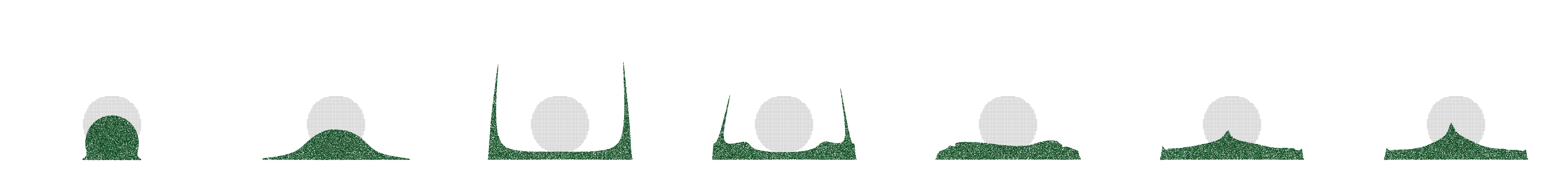}\\
\vspace{-1mm}
MLS-TLMPM: $t=\{0.14167,\:0.20833,\:0.47083,\:0.74167,\:1.31667,\:1.88750,\:2.06250\}$s\\
\includegraphics[width=\textwidth]{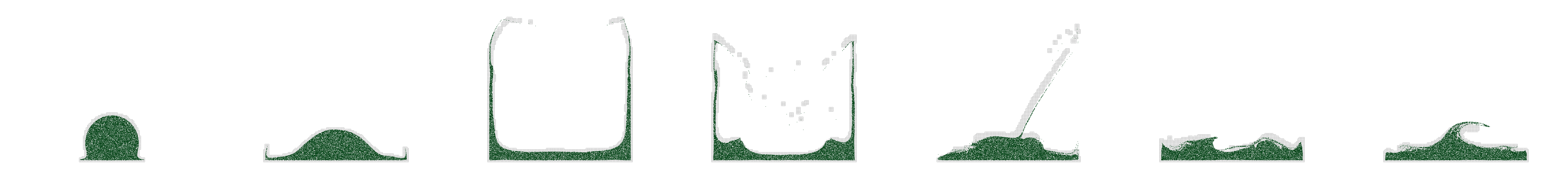}\\
\vspace{-1mm}
MLS-$A$-ULMPM: $t=\{0.14167,\:0.20833,\:0.47083,\:0.74167,\:1.31667,\:1.88750,\:2.06250\}$s\\
\includegraphics[width=\textwidth]{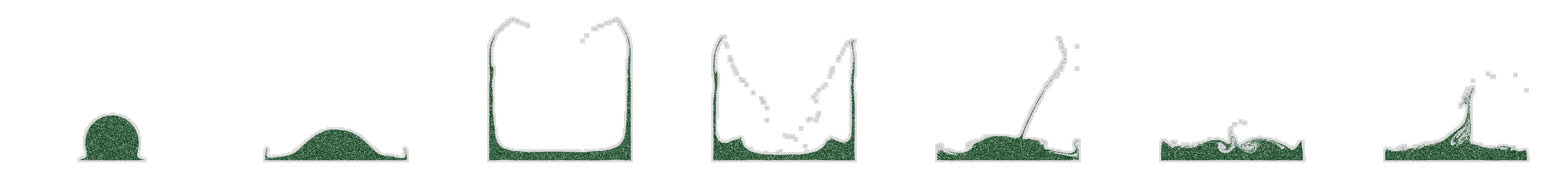}\\
\vspace{-2mm}
MLS-EMPM: $t=\{0.14167,\:0.20833,\:0.47083,\:0.74167,\:1.31667,\:1.88750,\:2.06250\}$s\\
\vspace{-4mm}
\end{center}
\caption{\textbf{2D Droplet.} 
We integrate $A$-ULMPM with traditional Kernel-MPM~\cite{Stomakhin:2013:MPMsnow} (see Appendix~\ref{sec:A-ULMPM-kernel}) and MLS-MPM (see Section~\ref{sec:A_UL_MLS_MPM}). 
Although TLMPM can eliminate numerical fracture in solid simulations as shown in Figure~\ref{fig:2D_rotating}, it fails to capture very large deformations, such as fluid-like motion (see rows 1 and 3) in both Kernel-EMPM and MLS-EMPM. Our proposed $A$-ULMPM automatically updates configurations to produce similarly detailed dynamics as those with EMPM for fluid simulation (see rows 4 and 5). Background grid represents the configuration linking to the present particle dynamics.}
\vspace*{-3mm}
\label{fig:2D_droplet}
\end{figure*}

 As a counterpart to Eulerian formulations, \emph{total Lagrangian formulations} offer a promising alternative for avoiding the cell-crossing instability and numerical fracture in MPM~\cite{De:2020:TLMPM,De:2021:TLMPM}. 
 Unlike EMPM, total Lagrangian MPM (TLMPM) measures stress and strain and computes derivatives and integrals with respect to the \emph{original} configuration at time $t^0$ (similar to traditional FEM~\cite{Sifakis:2012:FEM,Zienkiewicz:1977:FEM}). 
 By doing this, no matter what the deformation, the reference configuration being always the same, there is neither any cell-crossing instability nor any numerical fracture. 
 Despite its high efficacy in solid simulations, traditional TLMPM fails to model extremely large deformations that arise in fluid simulations. 
 We show a 2D droplet example in Figure~\ref{fig:2D_droplet}. TLMPM is not able to capture the dynamics of splashes because the interpolation kernel and its derivatives in TLMPM only reflect the fixed topology at time $t^0$ and do not support extreme topologically changing dynamics. 
%\subsection{Contributions}

To alleviate the cell-crossing instability and numerical fracture while allowing for large deformations, we present an arbitrary updated Lagrangian discretization of MPM ($A$-ULMPM) that spans from total Lagrangian formulations to Eulerian formulations. Unlike EMPM and TLMPM, $A$-ULMPM allows the configuration to be updated \emph{adaptively} for measuring physical states including stress, strain, interpolation kernels and their derivatives. 
As a natural extension, we revisit the APIC~\cite{Jiang:2017:APIC} and MLS-MPM~\cite{Hu:2018:Moving} formulations and integrate these two methods in $A$-ULMPM. We augment the accuracy of velocity rasterization by leveraging both the local velocity and its first-order derivatives. 
Our theoretical derivations focus on a nodal discretized Lagrangian (see equation~\eqref{eq:nodal_L}), instead of the weak form discretization in MLS-MPM~\cite{Hu:2018:Moving}, and naturally lead to a modified MLS-MPM in $A$-ULMPM, which can recover MLS-MPM by using a completely Eulerian formulation.  
$A$-ULMPM does not require significant changes to traditional EMPM, and is more computationally efficient since it only updates interpolation kernels and their derivatives when large topology changes occur. To summarize, our main contributions are as follows:
\vspace{-3mm}
\begin{enumerate}
    \item An arbitrary updated Lagrangian MPM ($A$-ULMPM) that spans discretizations from TLMPM to EMPM and avoids the cell-crossing instability and numerical fracture;
    \item An easy-to-implement criterion that automatically updates the reference topology to enable fluid-like simulations;
    \item Integration of APIC and MLS-MPM in $A$-ULMPM to allow angular momentum conservation and efficiency by reconstructing the interpolation kernel only during topology updates;
    \item End-to-end 3D simulations of stretching and twisting hyperelastic solids, splashing liquids, and multi-material interactions to highlight the benefits of our $A$-ULMPM framework.
\end{enumerate}
\section{RELATED WORK}
\label{sec:related-work}

In this work, we only review prior work related to MPM~\cite{jiang2016material} since our focus is on MPM. However, we note that there are several established methods for particle-based simulations, including SPH~\cite{Desbrun:1996:SPH}, position-based dynamics~\cite{Muller:2004:Point-based,muller:2007:pbd,macklin:2014:PBDs}, linear complementarity formulations~\cite{Erleben:2013:LCP}, and geometric computing techniques~\cite{deGoes:2015:PPIFS,Sin:2009:Voronoi}.

\subsection{MPM in Graphics}
The MPM discretization has been widely adopted in computer graphics applications~\cite{jiang2016material}. 
The seminal work of Zhu and Bridson~\cite{Zhu:2005:Animating} first introduced the FLIP method for sand simulation. Subsequent works further explored its strength in simulating a broader spectrum of material behaviors including snow~\cite{Stomakhin:2013:MPMsnow}, granular materials~\cite{Daviet:2016:Semi-implicit,Klar:2016:Drucker,Tampubolon:2017:Multi,gao2018animating}, foam~\cite{Ram:2015:Material,yue:2015:mpm-plasticity}, complex fluids~\cite{Fang:2019:Silly,Gao:2017:AGIMPM}, cloth, hair and fiber collisions~\cite{Jiang:2017:Anisotropic,Fei:2017:MSLHI,fei:2018:cloth}, fracture~\cite{Wolper:2019:CD-MPM,Wolper:2020:anisompm} and phase change~\cite{Stomakhin:2014:Augmented,gao2018gpu,Su:2021:USOSVL}. We also note the related works of~\cite{mcadams:2009:incomp} for hair simulation, \cite{sifakis:2008:incomp} for cloth simulation, \cite{narain:2010:sand} for sand simulation and~\cite{patkar:2013:bubble} for bubble simulation, which bear similarities to MPM due to their hybrid nature.
Various works have improved or modified aspects of the standard MPM techniques commonly used in graphics~\cite{Fang:2020:IQ-MPM,Yue:2018:HG,XUE:2020:NF, Ding:2019:Thermomechanical}.
Among them, notably, Jiang et al.~\shortcite{jiang:2015:apic,Jiang:2017:APIC} proposed an Affine Particle-In-Cell (APIC) approach that conserves angular momentum and prevents visual artifacts such as noise, instability, clumping and volume loss/gain existing in both FLIP and PIC methods. Furthermore, APIC was enhanced in~\cite{Fu:2017:PolyPIC,Hu:2018:Moving} to improve the kinetic energy conservation in particle/grid transfers. 
\begin{figure*}[h]
    \vspace{-4mm}
    \centering
    \includegraphics[width=\textwidth]{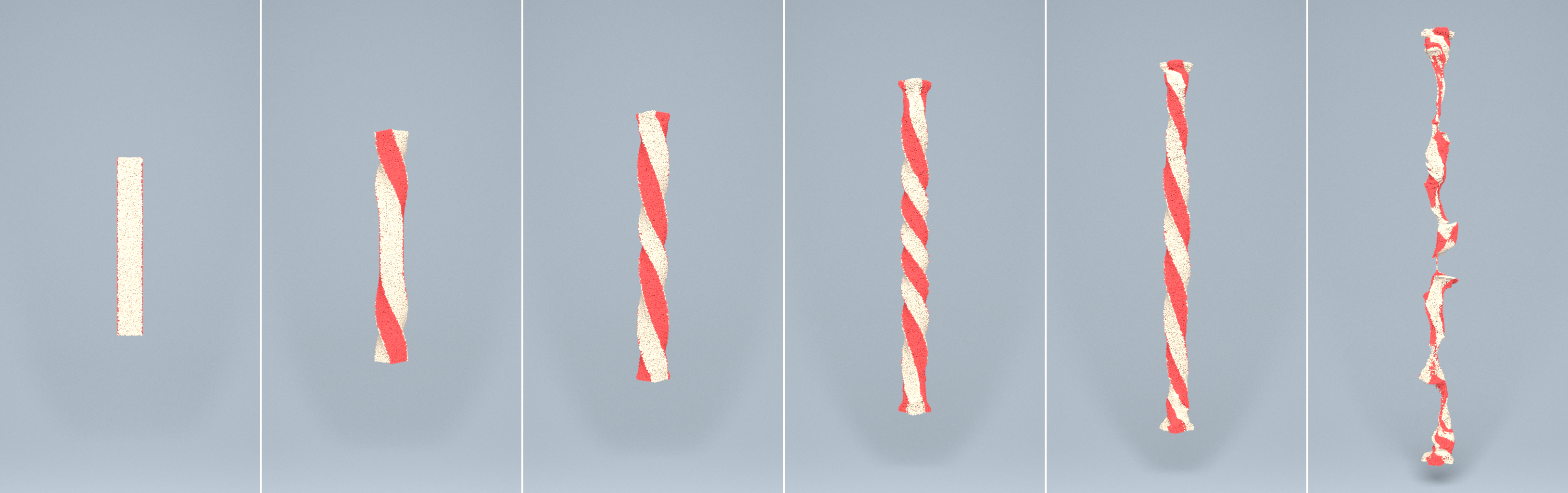}\\
     MLS-EMPM: Time instants from left to right $t=\{0,\:0.0005,\:0.0009,\:0.0016,\: 0.002,\:0.0027\}$s\\
    \includegraphics[width=\textwidth]{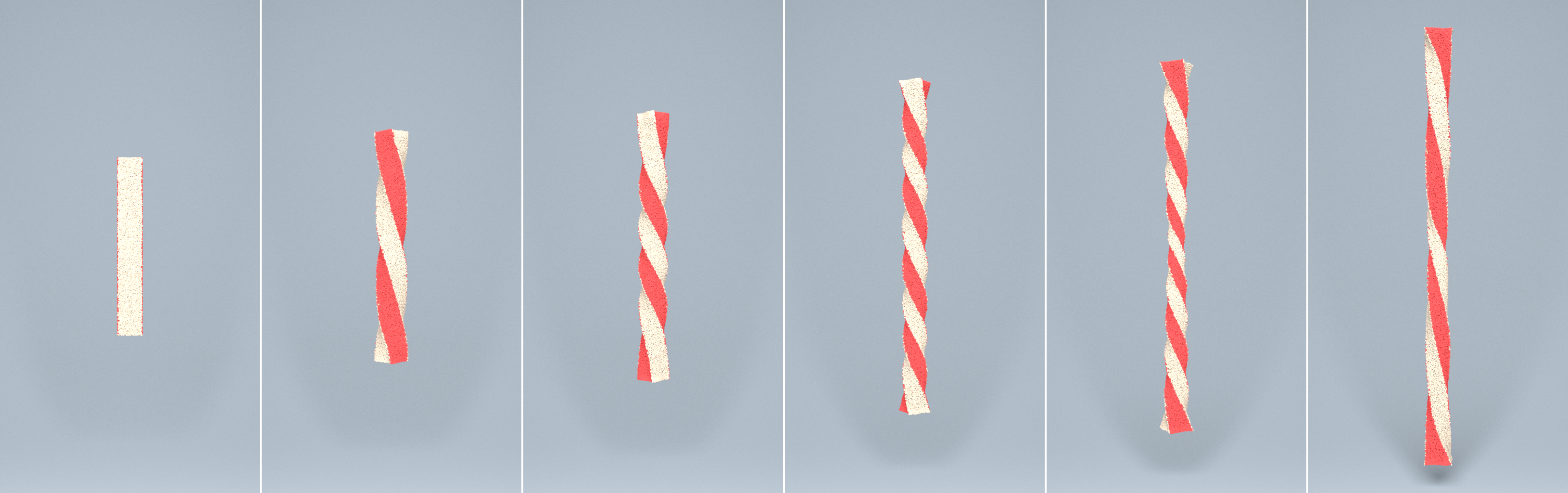}\\
     MLS-$A$-ULMPM:  Time instants from left to right $t=\{0,\:0.0005,\:0.0009,\:0.0016,\: 0.002,\:0.0027\}$s\\
     \vspace{-4mm}
    \caption{\textbf{3D Twisting column.} The two ends of a rectangular beam are kinematically separated while twisting the beam. Standard MLS-EMPM (top row) suffers from severe numerical fracture. In contrast, our MLS-$A$-ULMPM (bottom row) nicely captures the rich twisting surface and preserves column shape.}
    \label{fig:bar_twisting}
    \vspace{-3mm}
\end{figure*}

\subsection{MPM in Engineering}

In the engineering community, MPM was first introduced in~\cite{Sulsky:1994:Particle} as an extension of the FLIP method~\cite{Brackbill:1988:Flip}, and substantial improvements and variants have been proposed thereafter, including experimental validation for studying dynamic anticrack propagation in snow avalanches~\cite{gaume:2018:dynamic} and the use of MPM for designing differentiable physics engines for robotics applications~\cite{Hu:2019:ChainQueen}.
Different strategies for updating the stress were compared to investigate the energy conservation error in MPM in~\cite{Bardenhagen:2002:Energy}. The quadrature error and cell-crossing error of MPM was investigated in~\cite{Steffen:2008:Analysis}.
A generalized interpolation material point method (GIMPM)~\cite{Bardenhagen:2004:GIMPM} was proposed to obtain a smoother field representation by combining the shape functions of the grid with the particle characteristic function. 
The cell-crossing error in the MPM discretization was alleviated by introducing the local \emph{tangent} affine deformation of particles in the convected particle domain method (CPDI)~\cite{Sadeghirad:2011:CPDI}. 
However, CPDI does not completely remove the numerical fracture issue due to gaps between particles and grid cells. 
Recently, the standard MPM was reformulated with respect to the initial topology, which provides the total Lagrangian formulation for MPM (TLMPM)~\cite{De:2020:TLMPM}, which has been proven to completely eliminate the numerical fracture issue and cell-crossing instability and has been further extended in~\cite{De:2021:TLMPM} to support multi-body contacts.

%%%%====================================
\begin{comment}
\subsection{Moving Least Squares in MPM}
As a widely used fitting scheme, Moving Least Squares (MLS) has been utilized to boost the computational accuracy of various discretization methods~\cite{Lu:1994:newEFG, Zhang:2009:improvedEFG, Atluri:1998:New, Bessa:2014:RKPM, Zienkiewicz:1974:Least}. 
In the MPM discretization, MLS techniques have successfully demonstrated the ability of improving the solution accuracy in large deformation simulations~\cite{W:2007:improved, Sulsky:2016:improving}. 
However, the formulation requires the inversion of the moment matrix on the solution of a system of equations, therefore, this is both expensive and there is a possibility that the moment matrix may be singular or ill-conditioned.  
Spurious values due to the ill-conditioning of moment matrix were demonstrated in the background gird~\cite{W:2007:improved}. 
To circumvent this issue, Gram Schmidt orthogonalization method was used to formulate a set of orthogonal basis functions to facilitate rapid solution in the PolyPIC~\cite{Fu:2017:PolyPIC} and the Improved MPM~\cite{Tran:2019:Improved}.
This idea has also been used in the Element-free Galerkin method~\cite{Liew:2006:Boundary,Zhang:2009:improvedEFG,Zhang:2014:improved} to overcome the instability. 
Besides, a noteworthy integration of MLS and MPM was proposed in~\cite{Hu:2018:Moving}, which unifies APIC and PolyPIC and can be treated as a modified EFG method. 
\end{comment}
\section{Overview of Different Formulations for Continuum Mechanics}
\label{sec:governing_equation}
In this section, we briefly revisit the Lagrangian framework from continuum mechanics for describing the governing equations of motion and summarize the different formulations that can be derived depending on the choice of the reference configuration. 
%We then introduce an intermediate configuration at $t=t_s$ between the initial configuration $t=t_0$ and the current configuration $t=t_n$ as shown in Figure~\ref{fig:configrations}, and derive the associated governing equation at $t_s$ for continuum mechanics aiming to present  an arbitrary updated Lagrangian expression of continuum mechanics. 
\begin{figure}[h]
\centering
\begin{overpic}[width=.45\textwidth]{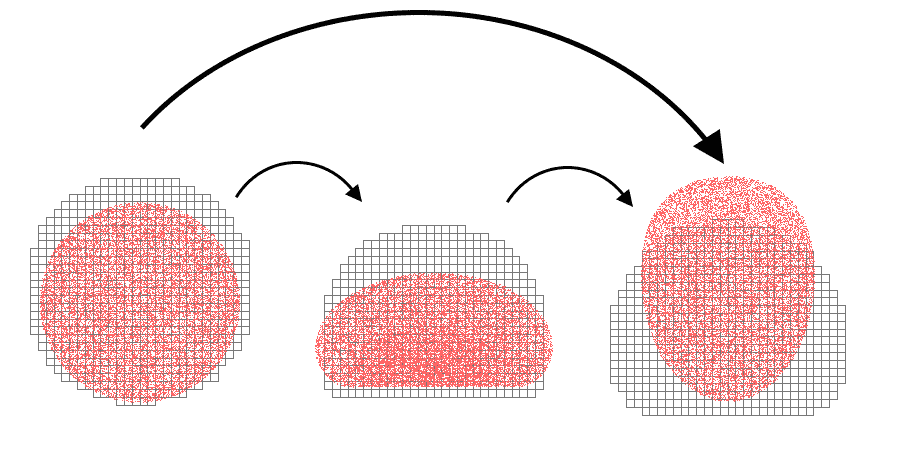}
    \put(8,0){$t=t_0$: $\bs{q}_0$}
    \put(42,0){$t=t_s$: $\bs{q}_s$}
     \put(75,0){$t=t_n$: $\bs{q}_n$}
    \put(25,35){$\bs{F}_{0s}=\frac{\p\bs{q}_s}{\p\bs{q}_0}$}
    \put(55,35){$\bs{F}_{sn}=\frac{\p\bs{q}_n}{\p\bs{q}_s}$}
    \put(40,43){$\bs{F}_{0n}=\frac{\p\bs{q}_n}{\p\bs{q}_0}$}
\end{overpic}
\vspace{-2mm}
\caption{\textbf{Elastic ball.} The deformation gradient can be described with respect to the initial configuration (total Lagrangian formulation) at time $t_0$ as $\bs{F}_{0s}$ and $\bs{F}_{0n}$, or with respect to an intermediate configuration at time $t_s$ (updated Lagrangian formulation) as $\bs{F}_{sn}$, where $\bs{q}$ is the displacement.}
\label{fig:configrations}
\vspace{-4mm}
\end{figure}
\subsection{Lagrangian Framework}
The Lagrangian $\f{L}$ for holonomic systems is defined as:
\begin{equation}
\f{L}(\bs{q},\dot{\bs{q}})=\f{K}(\bs{\dot{q}})-\f{U}(\bs{q}) %,\:\:\f{U}=\Pi_{int}+\Pi_{ext}
\end{equation}
where $\bs{q}$ is the generalized displacement, $\bs{\dot{q}}$ is the generalized velocity, $\f{K}(\dot{\bs{q}})$ is the kinetic energy and $\f{U}(\bs{q})$ is the potential energy. By omitting the energy due to external body forces and traction for simplicity, the kinetic and potential energies can be defined as:
\begin{equation}
\f{K}(\bs{\dot{q}})=\frac{1}{2}\int_{\f{B}}\bs{\dot{q}}^T\bs{\dot{q}} dV,\quad
\f{U}(\bs{q})=\int_{\f{B}}\rho_0\Psi(\bs{F})dV
\end{equation}
where $\rho_0$ is the material density at time $t_0$, 
$\Psi(\bs{F})$ denotes the Helmholtz free energy per unit mass in homogeneous materials, and $\bs{F}$ is the deformation gradient tensor. Consequently, the Lagrangian density function can be defined as follows:
\begin{equation}
     \bar{\f{L}}(\bs{q},\bs{\dot{q}}, \bs{F})=\frac{1}{2}\rho_0\bs{\dot{q}}^T\bs{\dot{q}} -\Psi(\bs{F})
\end{equation}
Based on the Lagrangian framework~\cite{Zienkiewicz:1977:FEM}, the governing equations of motion at time $t_n$ can be described as:
\begin{equation}
   \frac{d}{dt}\left(\frac{\p \bar{\f{L}}}{\p \bs{\dot{q}}_n}\right)-\frac{\p \bar{\f{L}}}{\p \bs{q}_n}=\bs{0}
    \label{eq:lagrangian}
\end{equation}
where
\begin{equation}
\frac{d}{dt}\left(\frac{\p \bar{\f{L}}}{\p \bs{\dot{q}}_n}\right)=\rho_0 \bs{\ddot{q}}_n, \quad \frac{\p \bar{\f{L}}}{\p \bs{q}_n}=\frac{\p\Psi(\bs{F})}{\p\bs{F}}\frac{\p\bs{F}}{\p \bs{q}_n}.%=\mathbb{P}\frac{\p\bs{F}}{\p \bs{q}},
\end{equation}
Note that the term ${\p\Psi(\bs{F})}/{\p\bs{F}}$ represents a stress tensor that is determined by the material constitutive model, and the term ${\p \bs{F}}/{\p\bs{q}}$ can be further expressed in terms of a divergence operator. These two terms together define the internal force as the material deforms. %We leave the $\frac{\p\Psi(\bs{F})}{\p\bs{F}}$ and  $\frac{\p\bs{F}}{\p \bs{q}}$ in equation~\eqref{}
%and $\mathbb{P}$ represents the \emph{first Piola–Kirchhoff stress}. 

\subsection{Total Lagrangian Formulation}

In this formulation, the stress and strain in the material are measured relative to the original configuration at time $t_0$ (see Figure~\ref{fig:configrations}), such that the deformation gradient tensor $\bs{F}$ is the displacement derivative of $\bs{q}_n$ with respect to $\bs{q}_0$. Substituting $\bs{F}_{0n}$ to $\p \bar{\f{L}}/\p \bs{q}_n$ gives:
 $$\frac{\p \bar{\f{L}}}{\p \bs{q}_n}=\frac{\p\Psi(\bs{F})}{\p\bs{F}_{0n}}\frac{\p}{\p \bs{q}_n}\left(\frac{\p \bs{q}_n}{\p \bs{q}_0}\right)=\nabla_0\cdot\mathbb{P}$$
 and have the following governing equation of motion:
    \begin{equation}
      \rho_0 \bs{\ddot{q}}_n=\nabla_{0}\cdot\mathbb{P}_0 
      \label{eq:Lagrangian GE}
    \end{equation}
where the divergence operator $\nabla_0$ is also evaluated with respect to the original configuration at time $t_0$. In the total Lagrangian formulation, the stress tensor $\mathbb{P}_0$ is the \emph{first Piola–Kirchhoff stress}. 

\subsection{Eulerian Formulation}

In this formulation, the stress and strain in the material are measured relative to the current configuration at time $t_n$ (see Figure~\ref{fig:configrations}). Thus, the expression for ${\p \bar{\f{L}}}/{\p \bs{q}}$ can be expanded as follows:
$$\frac{\p \bar{\f{L}}}{\p \bs{q}_n}=
\frac{\p\Psi(\bs{F})}{\p\bs{F}_{0n}}\frac{\p \bs{F}_{0n}}{\p \bs{q}_n}=
\frac{\p}{\p \bs{q}_n}\left(\frac{\p\Psi(\bs{F})}{\p\bs{F}_{0n}}\frac{\p \bs{q}_n}{\p \bs{q}_0}\right)=\nabla_n\cdot\left(\mathbb{P}_0\bs{F}_{0n}^T\right)$$
Besides, $\rho_0$ can be mapped to $\rho_n$ using the relation $\rho_0=J_{0n}\rho_n$, where $\rho_n$ is the density at time $t_n$ and $J_{0n}=\text{det}(\bs{F}_{0n})$. Consequently, the Eulerian formulation gives the following equation of motion:
\begin{equation}
    \rho_n\bs{\ddot{q}}=\nabla_n\cdot \mathbb{P}_n%{\left(\frac{1}{J_{0n}}\mathbb{P}_0\bs{F}_{0n}^T\right)}
    \label{eq:Euleraian GE}
\end{equation}
where $\mathbb{P}_n=\mathbb{P}_0\bs{F}_{0n}^T/{J_{0n}}$ provides the definition of the \emph{Cauchy stress}.

\subsection{Arbitrary Updated Lagrangian Formulation}
\label{sec:arbitrary-updated-lagrangian}

A general formulation for measuring stress and strain with respect to an arbitrary reference configuration at time $t_s$ has been derived in~\cite{Zienkiewicz:1977:FEM, Shabana:2018:computational}. In this formulation, the expression for ${\p \bar{\f{L}}}/{\p \bs{q}}$ can be expanded as follows:
%\begin{eqnarray}
%\frac{\p \bar{\f{L}}}{\p \bs{q}_n}=
%\frac{\p\Psi(\bs{F})}{\p\bs{F}_{0n}}\frac{\p \bs{F}_{0n}}{\p \bs{q}_n}=
%\frac{\p}{\p \bs{q}_s}\left(\frac{\p\Psi(\bs{F})}{\p\bs{F}_{0n}}\frac{\p \bs{q}_n}{\p \bs{q}_0}\frac{\p \bs{q}_s}{\p\bs{q}_n}\right)=\nabla_s\cdot\left(\mathbb{P}_0\bs{F}_{0s}^T\right)
%\label{eq:AUL}
%\end{eqnarray}
\begin{eqnarray}
\frac{\p \bar{\f{L}}}{\p \bs{q}_n}=
\frac{\p\Psi(\bs{F})}{\p\bs{F}_{0n}}\frac{\p \bs{F}_{0n}}{\p \bs{q}_n}=
\frac{\p\Psi(\bs{F})}{\p\bs{F}_{0n}}\frac{\p \bs{F}_{0n}}{\p \bs{F}_{sn}}\frac{\p \bs{F}_{sn}}{\p \bs{q}_n}=\nabla_s\cdot\left(\mathbb{P}_0\bs{F}_{0s}^T\right)
\label{eq:AUL}
\end{eqnarray}
Similar to the Eulerian formulation, we map $\rho_0$ to $\rho_s$ using the relation $\rho_s=J_{0s}\rho_s$, where $\rho_s$ represents the density at time $t_s$ and $J_{0s}=\text{det}(\bs{F}_{0s})$. Consequently, the arbitrary updated Lagrangian formulation gives the following governing equation of motion:
\begin{equation}
    \rho_s\bs{\ddot{q}}=\nabla_s\cdot\mathbb{P}_s%{\left(\frac{1}{J_{0s}}\mathbb{P}_0\bs{F}_{0s}^T\right)}
    \label{eq:generalized EG}
\end{equation}
where $\mathbb{P}_s=\mathbb{P}_0\bs{F}_{0s}^T/{J_{0s}}$ defines a stress measured at time $t_s$. 
By setting $s=0$ and using the defining properties of the initial configuration $J_{00}=1$ and $\bs{F}_{00}=\bs{I}$, where $\bs{I}$ is the identity matrix, equation~\eqref{eq:generalized EG} recovers the total Lagrangian formulation in equation~\eqref{eq:Lagrangian GE}. Likewise, setting $s=n$ yields the Eulerian formulation in equation~\eqref{eq:Euleraian GE}.

\begin{comment}
\tx{Known $\bs{\sigma}=p\bs{I}$, where $p$-pressure at $t_n$}
\begin{equation}
    \rho_n\bs{\ddot{q}}=\nabla_n\cdot\underbrace{\left(\frac{1}{J_{0n}}\mathbb{P}_0\bs{F}_{0n}^T\right)}_{\bs{\sigma}-in-MPM}
    \label{eq:Euleraian GE}
\end{equation}
\begin{enumerate}
\item Total Lagrangian
 \begin{equation}
      \rho_0 \bs{\ddot{q}}_n=\nabla_{0}\cdot\mathbb{P}_0 
      \label{eq:Lagrangian GE}
    \end{equation}
$$\sum_p V_p J_{0n}\bs{\sigma}\bs{F}_{on}^{-T} \nabla_0 W_{ij}^0$$

\item Updated Lagrangian--reference $t=t_s$
\begin{equation}
    \rho_s\bs{\ddot{q}}=\nabla_s\cdot{\left(\frac{1}{J_{0s}}\mathbb{P}_0\bs{F}_{0s}^T\right)}
    \label{eq:generalized EG}
\end{equation}
$$\frac{1}{J_{0s}}J_{0n}\bs{\sigma}\bs{F}_{on}^{-T}\bs{F}_{0s}^T$$
$$\frac{J_{0n}}{J_{0s}}\bs{\sigma}\frac{\p \bs{q}_0}{\p \bs{q}_n}\frac{\p \bs{q}_s}{\p \bs{q}_0}$$
$$\sum_p V_p \det(\bs{F}_{sn})\bs{\sigma}\underbrace{\frac{\p \bs{q}_n}{\p \bs{q}_s}^{-T}}_{\bs{F}_{sn}^{-T}}  \nabla_s W_{ij}^s$$

$$\bs{q}_{n+1}=\bs{q}_n+\bs{v}_n\dt$$
$$\frac{\p \bs{q}_{n+1}}{\p \bs{q}_s}=\frac{\p\bs{q}_s+\Delta\bs{q}}{\p \bs{q}_s}+\nabla_s\bs{v}_n\dt=\bs{F}_{sn}+\nabla_s\bs{v}_n\dt$$

$$\bs{F}_{s,n+1}=\frac{\p\bs{q}_n}{\p \bs{q}_s}+\nabla_s\bs{v}_n\dt=\bs{F}_{sn}+\nabla_s\bs{v}_n\dt$$
\item reference $t=0$
$$\bs{F}_{0,n+1}=\underbrace{(\bs{F}_{sn}+\nabla_s\bs{v}_n\dt)}_{\frac{\p\bs{q}_{n+1}}{\p\bs{q}_s}}\frac{\p \bs{q}_s}{\p \bs{q}_0}$$
\end{enumerate}
\end{comment}

%\input{tex_input/A_UL_nodal_discretization}
\section{Method Overview}
\label{sec:overview}
\begin{figure}[h]
%\vspace{-3mm}
\centering
\includegraphics[width=\columnwidth]{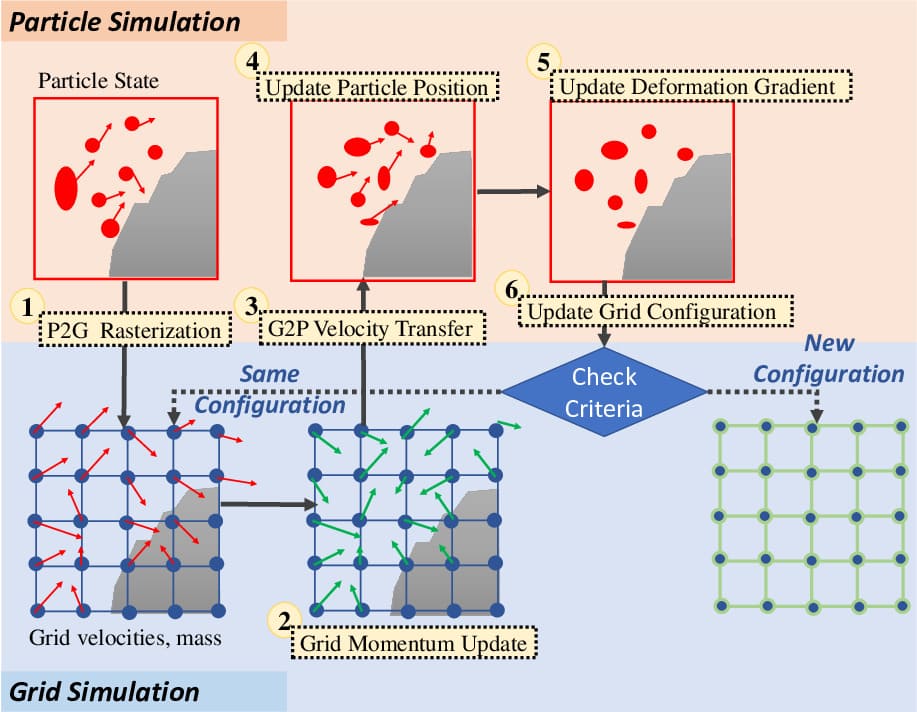}
\vspace{-7mm}
\caption{Overview of our proposed method.
The red and green arrows represent the velocity and force vectors at the nodes of the background grid. The blue grids represent the previous configuration while the green grids denote the new configuration according to our criterion for updating the state.}
\label{fig:A_ULMPM_overview}
\vspace{-3mm}
\end{figure}
\begin{figure*}[h]
    \vspace{-4mm}
    \centering
    \includegraphics[width=\textwidth]{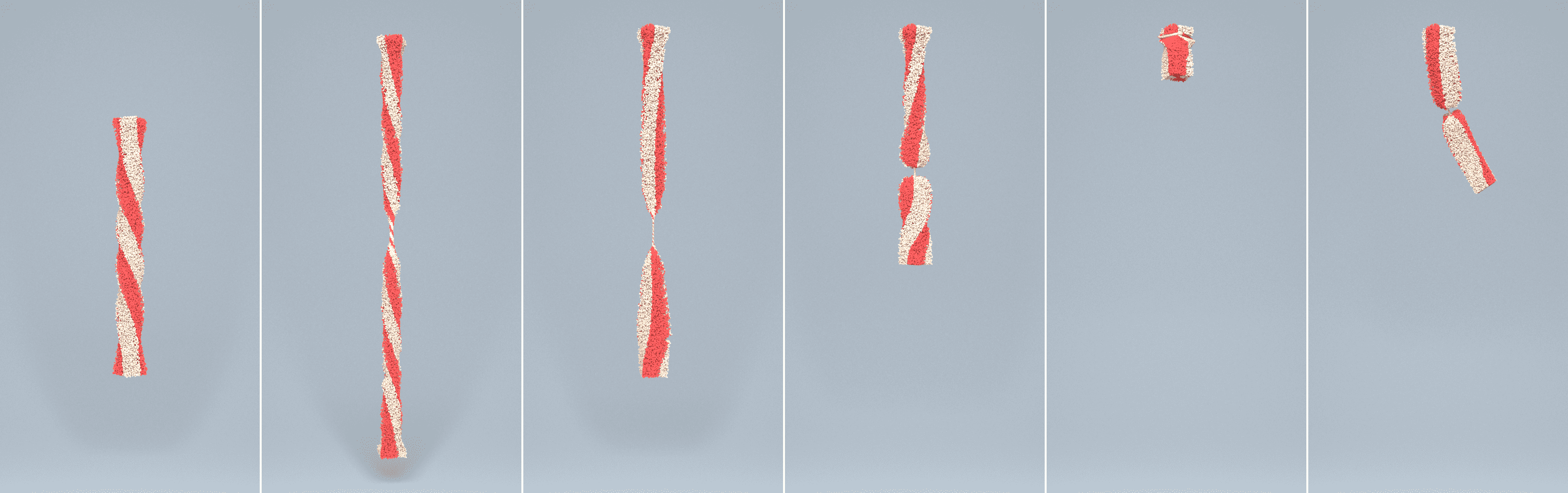} \\
   MLS-EMPM: $t=\{0.0008,\:0.0026,\:0.0032,\:0.0035,\: 0.04,\:0.1\}$s\\
    \includegraphics[width=\textwidth]{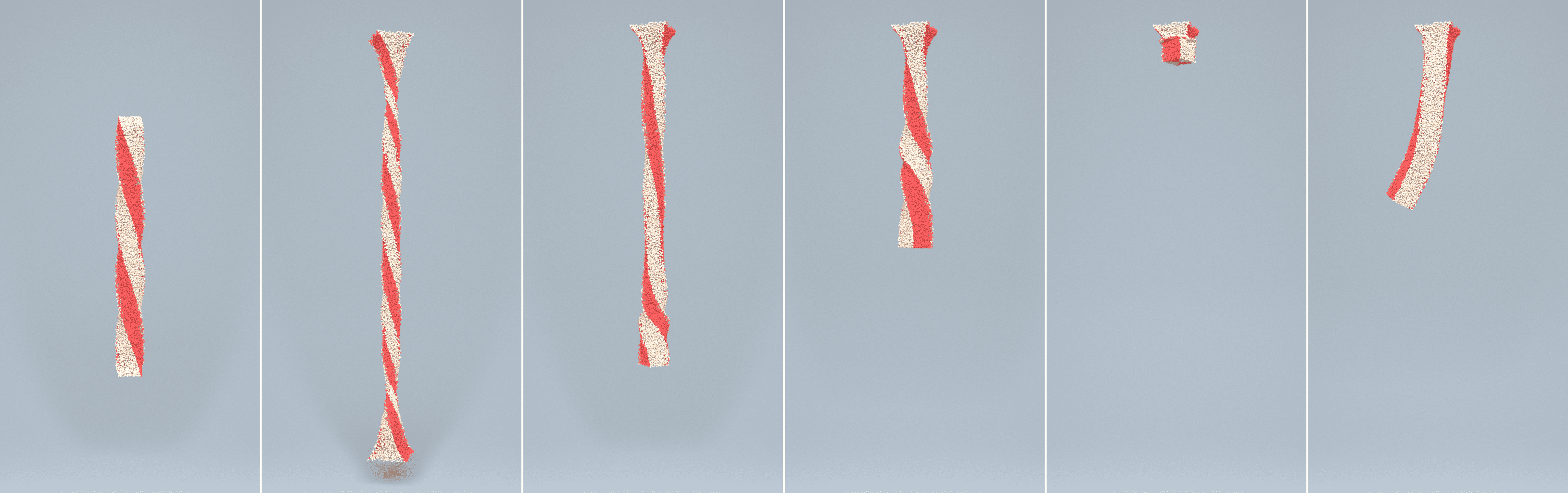}
    MLS-$A$-ULMPM: $t=\{0.0008,\:0.0026,\:0.0032,\:0.0035,\: 0.04,\:0.1\}$s\\
    \vspace{-4mm}
    \caption{The kinematic constraint on one end of the beam is released after a certain time. MLS-EMPM (row 1) fails to recover from the twisting deformation, while  MLS-$A$-ULMPM (row 2) produces realistic elastic response and recovers from the elastic deformation induced by the pulling and twisting motion.} 
    \label{fig:bar_twisting_inv_deformation}
    \vspace{-5mm}
\end{figure*}
Our method introduces a nodal discretized Lagrangian into hybrid grid-to-particle discretization for MPM and takes advantage of the zero cell-crossing error and zero numerical fracture properties of TLMPM. It also allows for updating the reference configuration in an adaptive manner to ensure robustness during extreme material deformations, such as those arising in fluid simulations. 
In contrast to prior MPM formulations (see equations~\eqref{eq:Lagrangian GE} and~\eqref{eq:Euleraian GE}), our method adopts adaptive reference configurations (see grids at time $t_s$ in Figure~\ref{fig:configrations}) to measure stress, strain, interpolation kernels and their derivatives. Data from particles is first transferred to grids (P2G) in their latest configuration. 
Next, the forces are evaluated and the velocities are updated on these grids. Subsequently, the updated grid velocities are interpolated back to the particles (G2P) and used to update the particle positions following the APIC method~\cite{jiang:2015:apic}. 
The velocity gradients and deformation gradients are then updated on the particles leveraging information on the grids. 
Our method uses a novel criterion (see equation~\eqref{eq:update_conf}) for automatically determining when to update the grid configuration. 
This prevents data transfers between the particles and grids from accumulating errors as the interpolation functions are only updated when necessary. Figure~\ref{fig:A_ULMPM_overview} provides a high-level overview of our method and the essential algorithmic steps are summarized below: % 
\begin{enumerate}
\item{\textbf{P2G Rasterization:}}
Velocities on particles are reconstructed by first-order Taylor expansion, and mass and momentum from particles at time $t_n$ are transferred to grids at time $t_s$. 
\item{\textbf{Grid Momentum Update:}} 
Grid momentum is updated using explicit or implicit schemes. Note that forces are evaluated with respect to the latest configuration at time $t_s$. 
\item{\textbf{G2P Velocity Transfer:}}
Use APIC~\cite{jiang:2015:apic} to transfer velocities from the grid to particles. 
\item{\textbf{Update Particle Positions:}}
Particle positions are updated with their new velocities.
\item{\textbf{Update Particle Deformation Gradients:}}
Use the MLS gradient operator to update the particle deformation gradient and account for plasticity, if it occurs.
\item{\textbf{Update Grid Configuration:}}
Check if the deformation is extreme according to the update criterion in equation~\eqref{eq:update_conf}. In case of a large deformation, update weights between particles and the grid and the $\mathbb{K}$ matrices on particles. 
\end{enumerate} 

\subsection{Terminology}
We show integration of our arbitrary updated Lagrangian formulation with Kernel-MPM~\cite{Stomakhin:2013:MPMsnow} in Appendix~\ref{sec:A-ULMPM-kernel} and MLS-MPM~\cite{Hu:2018:Moving} in Section~\ref{sec:A_UL_MLS_MPM}. In Kernel-MPM, the interpolation is performed using shape functions and stress divergence is evaluated by derivatives of the shape function, while MLS-MPM utilizes APIC particle-to-grid velocity rasterization and uses MLS shape functions to derive the internal force term. Using these definitions, our terminology for different methods is provided as follows:
\begin{enumerate}
    %\vspace{-1mm}
    \item {\textbf{Kernel-MPM}:}
    In kernel-MPM, we have kernel-EMPM and kernel-TLMPM represent the kernel-MPM in Eulerian and total Lagrangian frameworks, respectively. Kernel-EMPM was presented in~\cite{Stomakhin:2013:MPMsnow} and kernel-TLMPM was present in~\cite{De:2020:TLMPM}. These two methods can be recovered in our kernel-$A$-ULMPM framework by setting $s=n$ for EMPM and $s=0$ for TLMPM. 
    \item{\textbf{MLS-MPM}:}
    In MLS-MPM, we have MLS-EMPM and MLS-TLMPM represent MLS-MPM in Eulerian and total Lagrangian frameworks, respectively. 
    MLS-EMPM was presented in MLS-MPM~\cite{Hu:2018:Moving}.
    Besides, MLS-EMPM and MLS-TLMPM can be recovered in MLS-$A$-ULMPM by setting $s=n$ for MLS-EMPM and $s=0$ for MLS-TLMPM. 
\end{enumerate}
\begin{table}[h!]
\caption{Physical quantities stored on particles and grid nodes.}
\vspace{-3mm}
\begin{tabular}{ccc}
\hline
\textbf{Particle} & \textbf{Description} & \textbf{Grid}\\
\hline
$\bs{q}_p$ & position & $\bs{q}_i$ \\
$\bs{v}_p$ & velocity & $\bs{v}_i$ \\
\multirow{2}{*}{$\bs{F}_p^{0n}$} & deformation gradient at $t_n$ \\
                                & with respect to configuration at $t_0$&{$--$}\\
\multirow{2}{*}{$\bs{F}_p^{0s}$} & deformation gradient at $t_s$ \\
                                & with respect to configuration at $t_0$&$--$\\
\multirow{2}{*}{$\bs{F}_p^{sn}$} & deformation gradient at $t_n$ \\
                                & with respect to configuration at $t_s$&$--$\\
%$\bs{F}_p^{sn}$ & deformation gradient at $t_n$  w.r.t. Conf. at  $t_s$&$--$\\
$\mathbb{P}_p^{s}$ &Stress tensor &$--$\\
\multirow{2}{*}{$J_p^{sn}$}& volume change at $t_n$  \\
                           & with respect to configuration at $t_s$&$--$\\
$--$ & force & $\bs{f}_i$  \\
$V_p$ & volume & $--$ \\
$m_p$ &  mass & $m_i$ \\
%$\tl{\square}_p$& algorithmic qualities& $\tl{\square}_i$\\
\hline
\end{tabular}
\label{tab:notation}
\vspace{-7mm}
\end{table}
%\vspace{-4mm}
\section{$A$-ULMPM Formulation}
\label{sec:A_UL_MLS_MPM}

Starting from a nodal Lagrangian formulation, we derive an alternate expression for equation~\eqref{eq:Lagrangian GE} in the hybrid particle-grid framework of MPM. 
Interestingly, our formulation leads to a new MLS-MPM method (MLS-$A$-ULMPM) whose variant recovers MLS-MPM~\cite{Hu:2018:Moving} in the Eulerian setting.  
We use subscript $i$ to denote quantities on grid nodes, subscript $p$ to denote quantities on particles, and subscript $s$ to denote the intermediate configuration map. 
Table~\ref{tab:notation} summarizes the notation used in this section.
\begin{figure*}
    \centering
    \includegraphics[width=.98\textwidth]{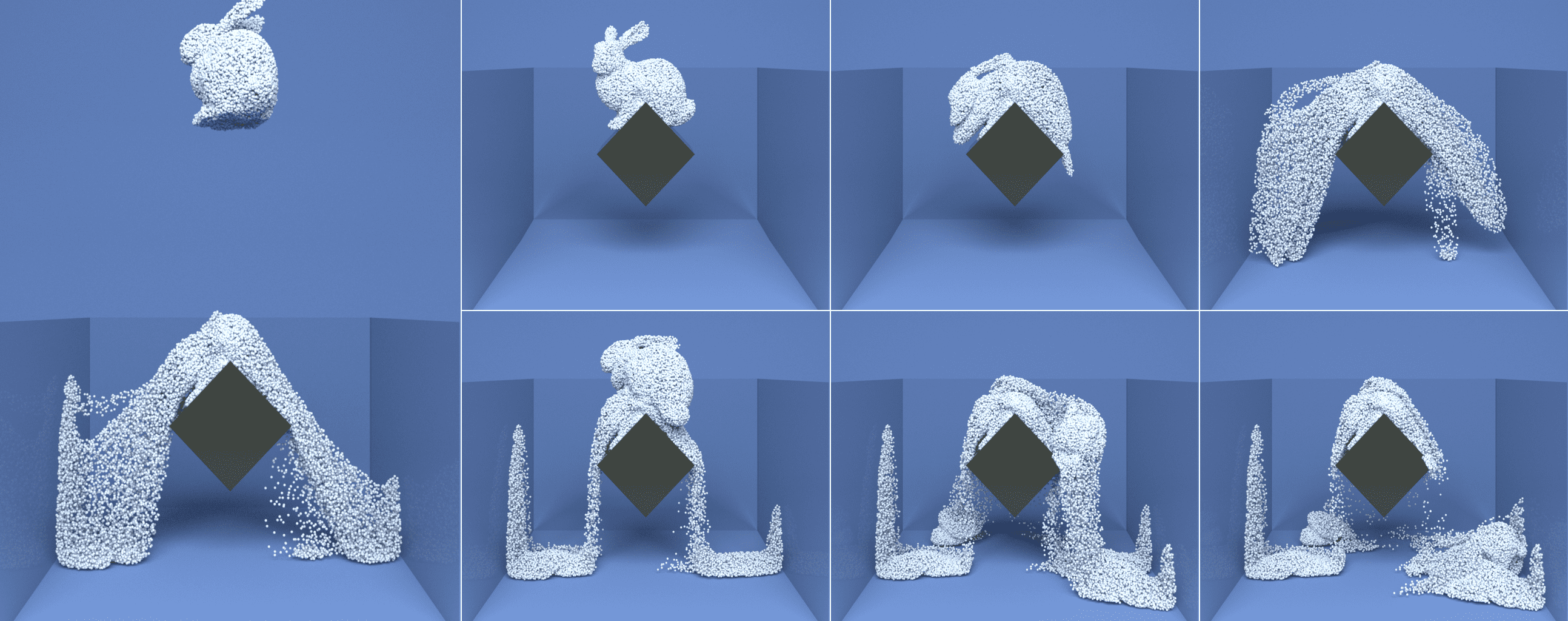}
     \vspace{-4mm}
     \caption{\textbf{Snow bunnies break over a wedge.} 
    Our $A$-ULMPM framework captures rich interactions of several snow bunnies smashing and scattering after falling on a solid wedge, demonstrating the extreme deformations that our method can capture, similar to its Eulerian counterparts proposed in prior works.} 
    \label{fig:snow}
    \vspace{-3mm}
\end{figure*}

\subsection{Grid Setting}

We use a global grid that covers the entire computational domain. Each object has its own \emph{configuration map} $\phi$ that maps points $\bs{x}$ within the object to specific locations $\phi(\bs{x})$ in the global grid. To reduce the associated memory overhead, we implement the global grid using sparsity aware data structures~\cite{Setaluri:2014:SSP}.%will be updated when criteria is satisfied (see section~\ref{sec:update_configuration}). 
%In practical simulations, global grid is only used for collision detecting. we embed all the individual configuration map to this global grid so that it can be directly integrate to any existing MPM. 
%========================

\subsection{P2G Rasterization}

In contrast to EMPM, the velocity gradients $\nabla_s\bs{v}_p$ and weights $W_{pi}^s$ are all evaluated with respect to the configuration map $\phi$ at time $t_s$. We use a first order Taylor expansion to reconstruct particle velocities and rasterize particle momentum to the grid as follows:
\begin{equation}
    m_i^s \bs{v}_i^n=\sum_p m_p \left(\bs{v}_p^n+ \nabla_s\bs{v}_p \bs{r}_{ip}^s\right) W_{pi}^s
\end{equation}
where $\bs{r}_{ip}^s=\left(\bs{q}_p^s-\bs{q}_i^s\right)$ and $\nabla_s \bs{v}_p^n=\frac{\p \bs{v}_p^n}{\p \bs{q}_s}$, which is further expanded out in equation~\eqref{eq:gradientVp}.
The mass/velocity at grid nodes are given by:
\begin{equation}
    m_i^s=\sum_p m_p W_{pi}^s,\quad \bs{v}_i^n= \frac{m_i^s \bs{v}_i^n} {\sum_p m_p W_{pi}^s}
\end{equation}
We also rasterize particle positions to the grid to simplify the theoretical derivations for the deformation gradient in equation~\eqref{eq:MPM_gradient_1} and internal forces in equation~\eqref{eq:EOM1} as shown below:
\begin{equation}
  \bs{q}_i^n=\frac{\sum_p\bs{q}_p^n W_{ij}^s}{\sum_p W_{ij}^s} 
 \label{eq:p2g_position}
\end{equation}
The idea of introducing a first-order Taylor expansion to enhance velocity rasterization is not new and can also be found in APIC~\cite{jiang:2015:apic} and MLS-MPM~\cite{Hu:2018:Moving}. 
%===========================================

\subsection{Grid Momentum Update} 
\subsubsection{Nodal Lagrangian}
We interpolate the Lagrangian $\f{L}_i^{n+1}$ at grid node $i$ using values  associated with its nearby particles $p$ as follows:
\begin{eqnarray}
    \f{L}_i^{n+1}=\sum_p \left[\frac{1}{2}\rho_j^0 \left(\bs{\dot{q}}_p^{n+1}\right)^T\bs{\dot{q}}^{n+1}_p -\Psi\left(\bs{F}_p^{0{n+1}}\right)\right]W_{ip}^s V_p^s
    \label{eq:nodal_L}
\end{eqnarray}
Substituting the expression for $\f{L}_i^{n+1}$ to equations~\eqref{eq:lagrangian} and~\eqref{eq:AUL} gives:
\begin{enumerate}
\item Kinetic term:
\begin{eqnarray}
     \frac{d}{dt}\left(\frac{\p \f{L}_i^{n+1}}{\p \bs{\dot{q}}_i^{n+1}}\right)=\sum_p \rho_p^0 \bs{\dot{q}}_p^{n+1} W_{ip}^s V_p^s =\sum_p \rho_p^0  W_{ip}^s V_p^s\bs{\ddot{q}}_p^{n+1}=m_i^s\bs{\ddot{q}}_p^{n+1}
\end{eqnarray}
\item Deformation term:
\begin{eqnarray}
    \frac{\p \f{L}_i^{n+1}}{\p \bs{q}_i^n}=-\sum_p \frac{\p \Psi(\bs{F}_p^{0{n+1}})}{\p\bs{q}_i^n} W_{ip}^s V_p^s= 
    -\sum_p\mathbb{P}_p^{s}  \frac{\p (\bs{F}_p^{s(n+1)})^T}{\p \bs{q}_i^s} W_{ip}^s V_p^s
\label{eq:defromation}
\end{eqnarray}
where $\mathbb{P}_j^s=\mathbb{P}_j^0(\bs{F}_j^{0s})^T/J_j^{0s}$. 
\end{enumerate}
\begin{figure*}[h]
    \centering
    \includegraphics[width=.98\textwidth]{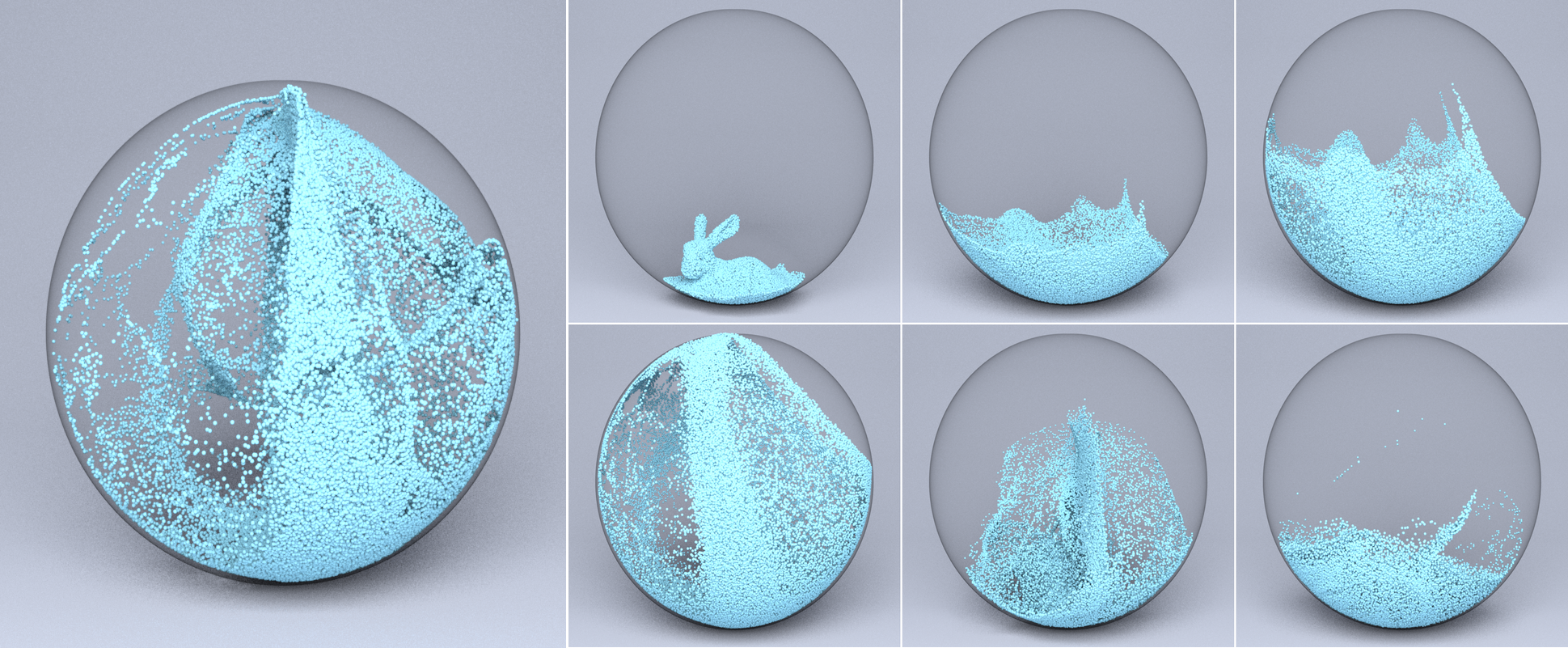}
     \vspace{-4mm}
     \caption{\textbf{Liquid bunny splash inside a ball.} A liquid bunny is dropped inside a spherical container and undergoes vibrant and dynamic splashes, demonstrating that our proposed $A$-ULMPM framework can capture the rich motion of incompressible fluids similar to existing Eulerian approaches proposed in prior works.}
    \label{fig:water}
    \vspace{-3mm}
\end{figure*}

Thus, the equation of motion for grid node $i$ at time $t_n$ is given by:
\begin{equation}
    m_i^s\bs{\ddot{q}}_i^{n+1}+\sum_j\mathbb{P}_p^s(\bs{F}_p^{0s})^T  \frac{\p (\bs{F}_p^{s(n+1)})^T}{\p \bs{q}_i^s} W_{ip}^s V_j^s=\bs{0}
    \label{eq:dis_L_eq}
\end{equation}
The reader may have noticed that an explicit expression of equation~\eqref{eq:dis_L_eq} relies on a concrete formulation of $\bs{F}_p^{s(n+1)}$ and its derivative ${\p (\bs{F}_p^{s(n+1)})}/{\p \bs{q}_i^s}$. 
For example, in kernel-MPM~\cite{Stomakhin:2013:MPMsnow}, the shape function is unitized to simplify the internal force evaluation in equation~\eqref{eq:dis_L_eq} (see Appendix~\ref{sec:A-ULMPM-kernel}). 

\subsubsection{MLS-based gradient operator}
We use the gradient operator based on \emph{moving least squares} (MLS) that was proposed in~\cite{Xue:2019:NL_L} (see Appendix~\ref{sec:mls_gradient}) that locally minimizes the error of a certain position over its neighborhood.  
Following the same notation in equation~\eqref{eq:dis_L_eq}, the deformation gradient of particle $p$ at time $t_n$ relative to an arbitrary configuration map at time $t_s$ is given as:
\begin{eqnarray}
    \bs{F}_p^{s(n+1)}=\left(\sum_{k}(\bs{q}_k^{n+1}-\bs{q}_p^{n+1})\otimes \bs{r}_{pk}^s W_{pk}^s\right)\left(\sum_{k}\bs{r}_{pk}^s\otimes\bs{r}_{pk}^s W_{pk}^s\right)^{-1}
\label{eq:gradient_qp_0s}
\end{eqnarray}
where $\bs{r}_{pk}^s=\bs{q}_k^s-\bs{q}_p^s$.
\subsubsection{Evaluation of $\p\bs{F}_p^{s(n+1)}/\p\bs{q}_i^s$}
Based on equation~\eqref{eq:dFdq} in Appendix~\ref{sec:mls_gradient}, we evaluate the derivative of $\bs{F}_p^{sn}$ with respect to $\bs{q}_i^s$ as: 
\begin{eqnarray}
\begin{small}
 \frac{\p \bs{F}_p^{s(n+1)}}{\p\bs{q}_i^s}=\left(\sum_{k}\left[\delta_{ik}^{n+1}-\delta_{ip}^{n+1}\right)\otimes \bs{r}_{pk}^s W_{pk}^s \right]\left[\sum_{k}\bs{r}_{pk}^s\otimes \bs{r}_{pk}^s W_{pk}^s V_k^s \right]^{-1}
 \end{small}
 \label{eq:dFjdqi}
\end{eqnarray}
\subsubsection{MPM Equation of Motion with Arbitrary Updated Lagrangian}
Substituting equation~\eqref{eq:dFjdqi} to equation~\eqref{eq:dis_L_eq} yields the following equation of motion for grid node $i$ at time $t_n$:
\begin{equation}
m_i^s\bs{\ddot{q}}_i^{n+1}+
\sum_p\left[\mathbb{P}_p^s\mathbb{K}_p^s+\mathbb{P}_i^s\mathbb{K}_i^s\right]\bs{r}_{ip}^s W_{ip}^s V_p^s=\bs{0}
\label{eq:EOM1}
\vspace{-3mm}
\end{equation}
where $\mathbb{P}_p^s=\mathbb{P}_p^0(\bs{F}_p^{s(n+1)})^T$ and $\mathbb{P}_i^s=\mathbb{P}_i^0(\bs{F}_i^{s(n+1)})^T$. $\mathbb{K}_p^s$ and $\mathbb{K}_i^s$ are given by:
\begin{equation*}
    \mathbb{K}_p^s=\left(\sum_i \bs{r}_{pi}^s\otimes\bs{r}_{pi}^s W_{pi}^s V_i^s\right)^{-1},\quad \mathbb{K}_i^s=\left(\sum_p \bs{r}_{ip}^s\otimes\bs{r}_{ip}^s W_{ip}^s V_p^s\right)^{-1}. 
\end{equation*}
We take advantage of equation~\eqref{eq:p2g_position} to simplify the term $\mathbb{P}_i^0(\bs{F}_i^{0s})^T\mathbb{K}_i^s$ in equation~\eqref{eq:EOM1} as follows:
$$\sum_p\mathbb{P}_i^s\mathbb{K}_i^s\bs{r}_{ip}^sW_{ip}^sV_p^s=\mathbb{P}_i^s\mathbb{K}_i^s\sum_p\bs{r}_{ip}^sW_{ip}^sV_p^s=0$$ %\left({\sum_p \bs{q}_p^s W_{ip}^s}-{\sum_p \bs{q}_i^sW_{ip}^s}\right)V_p^0=\bs{0}.$$
Setting $V_p^s=J_p^sV_p^0$, equation~\eqref{eq:EOM1} can be rewritten as follows:
\begin{eqnarray}
m_i^s\bs{\ddot{q}}_i^{n+1}+
\sum_p \mathbb{P}_p^0(\bs{F}_p^{0s})^T\mathbb{K}_p^s\bs{r}_{ip}^s W_{ip}^s V_p^0=\bs{0}
\label{eq:EOM2}
\vspace{-3mm}
\end{eqnarray}
Equation~\eqref{eq:EOM2} is a general MPM formulation that allows the use of arbitrary intermediate configurations. It spans from total Lagrangian formulations to updated Lagrangian formulations. Specifically:
\begin{enumerate}
    \item Setting $s=0$ yields total Lagrangian MPM:
\begin{eqnarray}
m_i^0\bs{\ddot{q}}_i^{n+1}+
\sum_p \mathbb{P}_p^0\mathbb{K}_p^0\bs{r}_{ip}^0 W_{ip}^0 V_p^0=\bs{0}
\label{eq:TLMPM}
\end{eqnarray}

\item Setting $s=n+1$ yields Eulerian MPM:

\begin{eqnarray}
m_i^n\bs{\ddot{q}}_i^{n+1}+
\sum_p \mathbb{P}_p^0(\bs{F}_p^{0(n+1)})^T\mathbb{K}_p^{n+1}\bs{r}_{ip}^{n+1} W_{ip}^{n+1} V_p^0=\bs{0}
\label{eq:EMPM}
\end{eqnarray}
\end{enumerate}
In the total Lagrangian formulation, there is no need to update $W_{ip}$, $\bs{r}_{ip}$, $m_i$, and $\mathbb{K}_p$ matrices in equation~\eqref{eq:TLMPM}, while they require an update at every time step in the Eulerian setting (see equation~\eqref{eq:EMPM}). 
\begin{figure*}[h!]
    \centering
    \includegraphics[width=\textwidth]{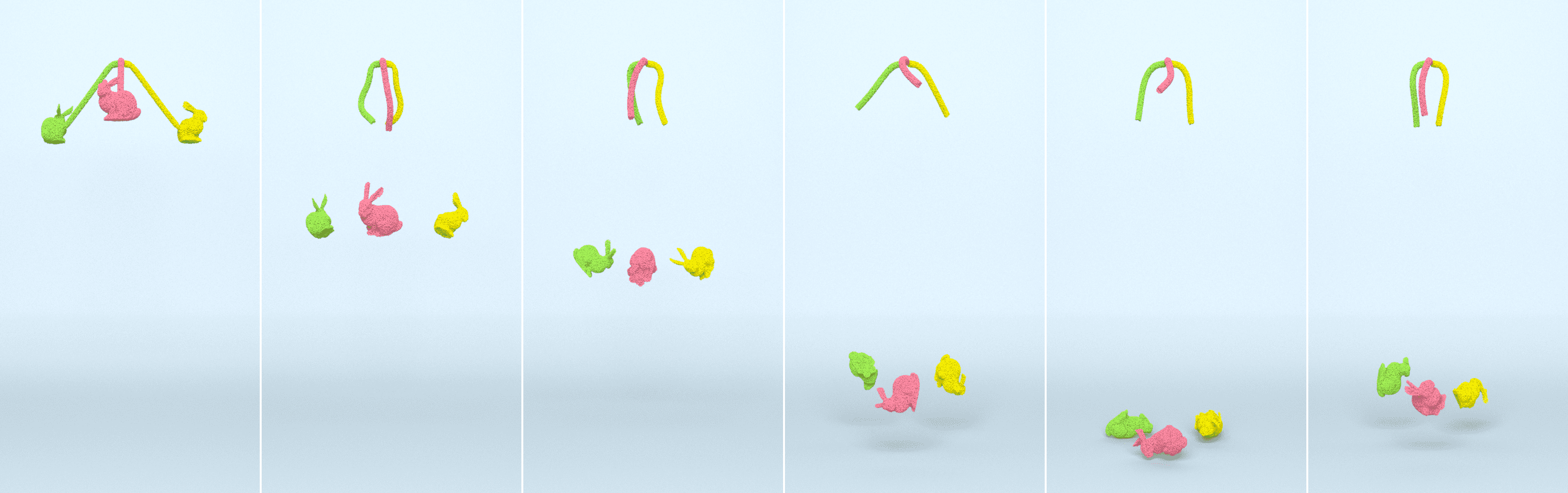}\\
    \includegraphics[width=\textwidth]{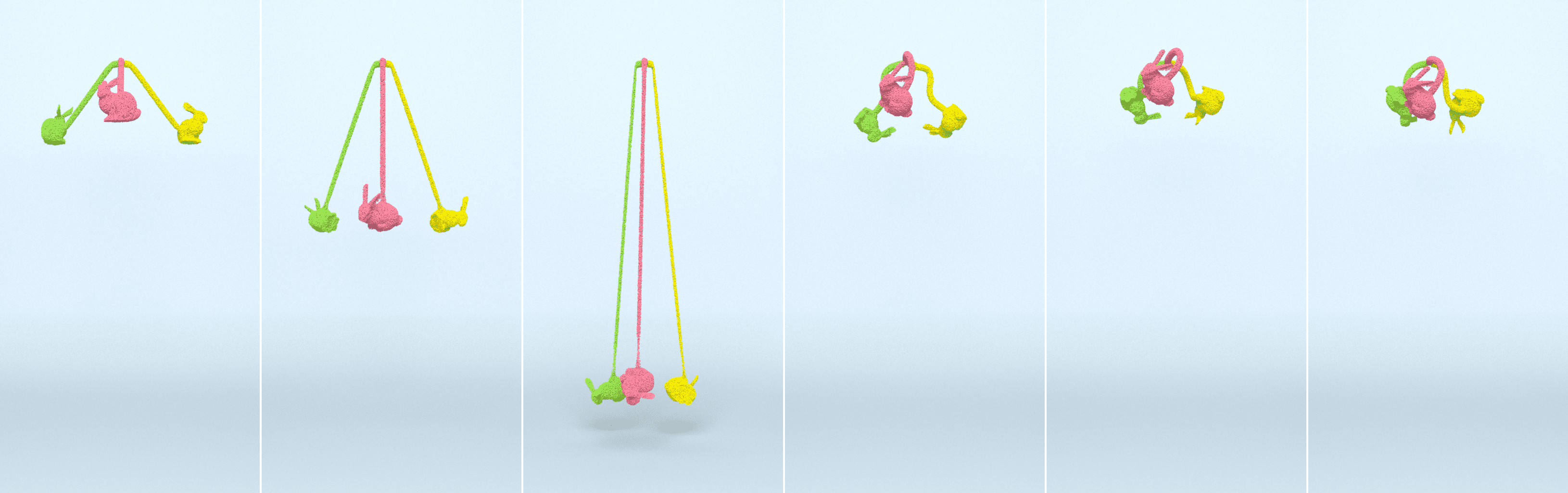}\\
     \vspace{-4mm}
     \caption{\textbf{Hyperelastic bunny yo-yo.} Under the effects of gravity, a hyperelastic bunny yo-yo breaks mid-way due to numerical fracture, when simulated with MLS-EMPM (top), while MLS-$A$-ULMPM (bottom) robustly captures the stretching motion of the elastic cord to pull the bunnies back upwards.}
    \label{fig:strechy_bunny_1}
        \vspace{-3mm}
\end{figure*}
\vspace{-3mm}
\subsection{Grid Velocity and Position Update}
With explicit time stepping, the velocity is updated as $\bs{v}_i^{n+1}=\bs{\hat{v}}_i^{n+1}$, where $\bs{\hat{v}}_i^{n+1}$ is given by: 
\begin{equation}
\bs{\hat{v}}_i^{n+1}=\bs{v}_i^n-\frac{\dt}{m_i}\sum_p\mathbb{P}_p^0(\bs{F}_p^{0s})^T\mathbb{K}_p^s\bs{r}_{ip}^s W_{ip}^s V_j^0\\
\label{eq:grid_Vel_update_explicit}
\end{equation}
For a semi-implicit update, we follow~\cite{Stomakhin:2013:MPMsnow, Hu:2018:Moving} and take an implicit step on the velocity update by utilizing the Hessian of $\f{L}^{n+1}$ with respective to $\bs{q}_i^{n+1}$. 
The action of this Hessian on an arbitrary increment $\delta \bs{q}^s$ is given as follows:
 \begin{equation}
     -\delta \bs{f}_i=\sum_pV_p^0 \mathbb{A}_p^s(\bs{F}_p^{0s})^T\mathbb{K}_p^s\bs{r}_{ip}^s W_{ip}^s V_j^0
     \label{eq:Hessian_L_derivative}
 \end{equation}
 where $\mathbb{A}_p^s$ is given by: 
 \begin{equation}
     \mathbb{A}_p^s=\frac{\p ^2 \f{L}^{n+1}}{\p \bs{F}_p^{0s} \p\bs{F}_p^{0s}}:\sum_j\delta \bs{q}_j^s \mathbb{K}_p^s\bs{r}_{ip}^s W_{ip}^s V_j^0 \bs{F}_p^{0s}.
 \end{equation}
 We linearize the implicit system with one step of Newton's method, which provides the following symmetric system for $\bs{\hat{v}}_i^{n+1}$:
 \begin{equation}
 \sum_j\left(\bs{I}\delta_{ij}+\frac{\dt^2}{m_i}\frac{\p ^2 \f{L}^{n+1}}{\p \bs{q}_i^s \p\bs{q}_j^s}\right) \bs{v}_j^{n+1}=\bs{\hat{v}}_i^{n+1}
 %\sum_j\left(\bs{I}\delta_{ij}-\frac{\dt^2}{m_i}\frac{\p \bs{f}_i^n}{ \p\bs{q}_j}\right)\bs{v}_j^{n+1}=\bs{\hat{v}}_i^{n+1}
 \label{eq:grid_Vel_update_implicit}
 \end{equation}
 where $\bs{I}$ is the identity matrix and $\bs{\hat{v}}_i^{n+1}$ is given in equation~\eqref{eq:grid_Vel_update_explicit}. We update rasterized positions at grid nodes as shown below:
 \begin{equation}
  \bs{q}_i^{n+1}=\bs{q}_i^n+\dt\bs{{v}}_i^{n+1}
 \end{equation}

\subsection{G2P Velocity and Position Transfer} 
We project $\bs{q}_i^{n+1}$ to the global grid and process grid-based collisions~\cite{Stomakhin:2013:MPMsnow} to compute $\bs{v}_i^{n+1}$, which is transferred back to particles using the APIC method~\cite{jiang:2015:apic}. %More specifics can be found in~\cite{Jiang:2017:APIC}. 
The specific updates for $\bs{q}_p^{n+1}$ and $\bs{v}_p^{n+1}$ are given below:
\begin{equation}
    \bs{v}_p^{n+1}=\sum_i\bs{v}_i^{n+1}W_{pi}^s,\quad \bs{q}_p^{n+1}=\bs{q}_p^n+\dt \bs{v}_p^{n+1}
\end{equation}
\subsection{Update Particle Deformation Gradients}
We first update particle deformation gradients with respect to the latest configuration at time $t_s$ and use the following MLS-based gradient operator (see equation~\eqref{eq:gradient_qp_0s}):
\begin{equation}
    \bs{F}_p^{s(n+1)}=\frac{\p\bs{q}^{n+1}_p}{\p\bs{q}^s_p}=
    \left(\sum_{i}(\bs{q}_i^{n+1}-\bs{q}_p^{n+1})\otimes \bs{r}_{ip}^s W_{pi}^s V_i^s\right)\mathbb{K}_p^s
\label{eq:MPM_gradient_1}
\end{equation}
Substituting $\bs{q}_i^{n+1}=\bs{q}_i^n+\dt\bs{v}_i^{n+1}$ and $\bs{q}_p^{n+1}=\bs{q}_i^n+\dt\bs{v}_p^{n+1}$ to equation~\eqref{eq:MPM_gradient_1} gives:
\begin{eqnarray}
\begin{split}
  \bs{F}_p^{s(n+1)}=  &\sum_i\left[\left(\bs{q}_i^n+\bs{v}_i^{n+1}\dt\right)-\left(\bs{q}_p^n+\dt\bs{v}_p^{n+1}\right)\right]\otimes \bs{r}_{pi}^s W_{pi}^s V_i^s \mathbb{K}_p^s\\
   % &=\sum_i\left(\bs{q}_i^n-\bs{q}_p^n\right)\otimes \bs{r}_{pi}^s W_{pi}^s V_i^s \mathbb{K}_p^s+\dt\sum_i\left(\bs{v}_i^{n+1}-\bs{v}_p^{n+1}\right)\otimes \bs{r}_{pi}^s W_{pi}^s V_i^s \mathbb{K}_p^s \\
   =  &\bs{F}_p^{sn}+\dt\nabla_s\bs{v}_p^{n+1}\\
\end{split}
\label{eq:MPM_gradient_s(n+1)}
\end{eqnarray}
where $\nabla_s\bs{v}_p^{n+1}$ is given by:
\begin{equation}
\nabla_s\bs{v}_p^{n+1}=\sum_i(\bs{v}_i^{n+1}-\bs{v}_p^{n+1})\otimes \bs{r}_{pi}^s W_{pi}^s V_i^s \mathbb{K}_p^s
\label{eq:gradientVp}
\end{equation}
Using the chain rule, the update for $\bs{F}_p^{0(n+1)}$ is given by:
\begin{equation}
    \bs{F}_p^{0(n+1)}=\frac{\p \bs{q}_p^{n+1}}{\p\bs{q}_p^{s}}\frac{\p\bs{q}_p^{s}}{\p\bs{q}_p^{0}}
    =\bs{F}_p^{s(n+1)}\bs{F}_p^{0s}
\end{equation}

\begin{figure*}
    \centering
    \includegraphics[width=\textwidth]{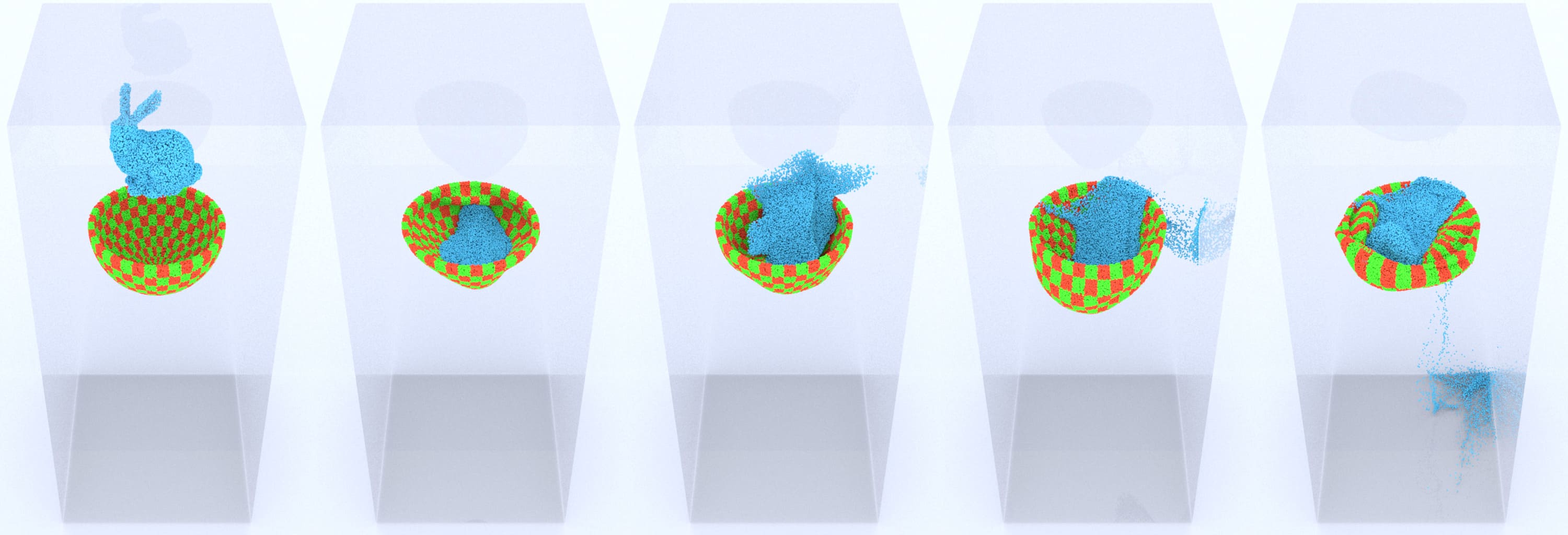}
    \includegraphics[width=\textwidth]{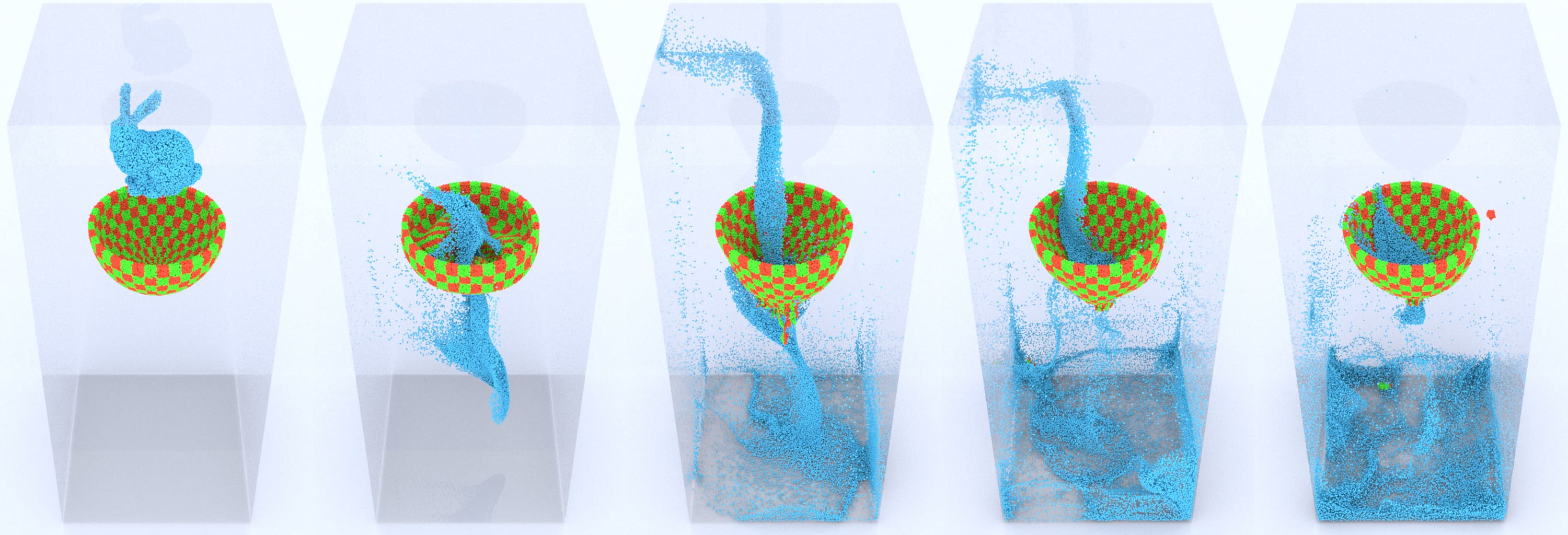}
    \vspace{-7.5mm}
    \caption{\textbf{Liquid bunny falling on a hyperelastic bowl.} (Top) Our proposed $A$-ULMPM framework can robustly capture the vivid dynamic responses of fluid-solid interactions and preserves the bowl shape. (Bottom) In contrast, the bowl fractures and fails to hold the water when simulated with MLS-EMPM.}.
    \label{fig:FSI}
    \vspace{-6mm}
\end{figure*}

\subsection{Update Grid Configuration Map}\label{sec:update_configuration}
We designed the configuration update criterion based on an intuitive assumption that more severe deformation is accompanied with large change of $J$ in the next time step $t^{n+1}$ with respect to the latest configuration $t^s$. 
Using this observation, our criterion is defined as a measure for the amount of particle deformation as follows:
\begin{equation}
    \delta J_p=\|J_p^{s(n+1)}-J_p^{ss}\|
    \label{eq:update_conf}
\end{equation}
where $J_p^{s{n+1}}=\text{det}(\bs{F}_p^{s(n+1)})$ and $J_p^{ss}=1$. 
We mark particles with $\delta J_p\geq \epsilon$ as indicators where large deformation is taking place and count the total amount of marked particles ($n_{mp}$). If $n_{mp}/n_p\geq \eta$, where $n_p$ is the total number of particles, we update $W_{ip}^s$, $\mathbb{K}_p^s$, and $\bs{F}_p^{0s}$. Otherwise, these variables remain the same until a new update occurs. $\epsilon$ and $\eta$ are user-defined parameters to adjust the update frequency. 
%{For solid simulations, we set $\epsilon=0.5$ and $\eta=0.5$, while for fluid simulations, we set $\epsilon=0.1$ and $\eta=0.01$.}
\subsection{Similarities with  MLS-EMPM and APIC}
While our $A$-ULMPM framework is new to computer graphics, setting $s=n$ for the configuration map shares some similarities with MLS-EMPM~\cite{Hu:2018:Moving} and APIC~\cite{jiang:2015:apic}. 
We prove that the MLS-gradient of velocity is equivalent to the APIC-gradient of velocity by leveraging the properties of interpolation weights $\sum_i \bs{r}_{pi} W_{pi} =\bs{0}$~\cite{Jiang:2017:APIC} as follows:
\begin{eqnarray}
\begin{split}
    \nabla\bs{v}_p=&\sum_i(\bs{v}_i-\bs{v}_p)\otimes\bs{r}_{pi} W_{ip}V_i
    \mathbb{K}_p\\
    =&\underbrace{\sum_i\bs{v}_i\otimes\bs{r}_{pi} W_{ip}V_i
    \mathbb{K}_p}_{\textbf{APIC:} \nabla \bs{v}_p}-\underbrace{\bs{v}_p\otimes\sum_i\bs{r}_{pi} W_{ip}V_i
    \mathbb{K}_p}_{\bs{0}}
\end{split}
\end{eqnarray}
Our internal force in equation~\eqref{eq:grid_Vel_update_explicit} has the same pattern as that in original MLS-EMPM~\cite{Hu:2018:Moving}, which is derived from the weak-form element-free Galerkin (EFG) framework. The $\mathbb{K}_p$ matrices are simplified by $\frac{4}{\dx^2}\bs{I}$ for quadratic $W_{ip}$ and $\frac{3}{\dx^2}\bs{I}$ for cubic $W_{ip}$. This simplification comes from the properties of splines, and is not generally true for other interpolation functions. 
\vspace{-4mm}

%\tx{Only when particles are located at cell center, our $\mathbb{K}_p$ recovers the above two  cases.}

\section{Results}
\label{sec:result}

Accompanying this article, we open-source our code for running 3D examples with our proposed $A$-ULMPM framework (see Section~\ref{sec:A_UL_MLS_MPM}). MLS-based MPM, including MLS-EMPM, MLS-TLMPM, and MLS-$A$-ULMPM, was applied to simulate all our 3D simulations. 
Besides the advantage of removing numerical fracture and the cell-crossing instability in solid and fluid simulations, $A$-ULMPM has the practical convenience of using the same numerical implementation to adaptively update the configuration map, without having to switch between TLMPM and EMPM.
We use the example of a falling elastic ball (see Figure~\ref{fig:configrations}) to evaluate the computational cost of explicit (see equation~\eqref{eq:grid_Vel_update_explicit}) and implicit (see equation~\eqref{eq:grid_Vel_update_implicit}) Euler schemes. 
As shown in the first two rows of Table~\ref{tab:timings}, the computational cost of these two schemes are comparable. 
Given the low stiffness in the materials, we utilize the explicit Euler scheme in all our 3D examples and did not experience any need for excessively small time steps.  
For our 3D solid simulations, we set $\epsilon=0.5$ and $\eta=0.1$, and observed that no configuration update was required, showing that $A$-ULMPM inherits the advantage of TLMPM for solid simulation. 
Due to foreseeable large deformations that arise when simulating fluid-like materials, such as water splashes and snow scattering, we set $\epsilon=0.01$ and $\eta=0.01$ to update the configuration maps more frequently. 
Even so, our $A$-ULMPM scheme reduces the configuration update overhead by $4.27\times$ -- $31.52\times$ in fluid-like simulations (see Table~\ref{tab:timings2}), compared to  standard EMPM. 
Table~\ref{tab:timings} summarizes the specific timings for all our examples. 

\begin{table}[h!]
\vspace{-3mm}
\caption{All simulations were run on Machine 1: Intel(R) Core(TM) i7-8750H CPU @ 2.20GHz and Machine 2: Intel(R) Xeon(R) CPU E5-1620 v4 @ 3.50GHz. {Simulation time is measured in average seconds per time step}. \textbf{Grid:} The number of occupied voxels in the background sparse grid. \textbf{Particle:} The total number of MPM particles in the simulation.}
\vspace{-4mm}
\begin{center}
\setlength{\tabcolsep}{0.1mm}{
\begin{tabular}{lcccc}
\hline 
\textbf{Simulation} & \textbf{Time} &  \textbf{Machine} &  \textbf{Grid} & \textbf{Particle} \\
\hline 
2D Elastic ball (Fig.\ref{fig:configrations})           & 0.0148    & 1     & 16.4K     & 20K      \\
2D Elastic ball (implicit)                              & 0.0165    & 1     & 16.4K     & 20K     \\
2D Rotation (Fig.\ref{fig:2D_rotating}(1st row))        & 0.0571    & 1     & 16.4K     & 42K    \\
2D Rotation (Fig.~\ref{fig:2D_rotating}(2nd row))       & 0.0571    & 1     & 16.4K     & 42K      \\
2D Rotation (Fig.~\ref{fig:2D_rotating}(3rd row))       & 0.0571    & 1     & 16.4K     & 42K     \\
2D Droplet (Fig.~\ref{fig:2D_droplet}(1st row))  & 0.0154    & 1     & 16.4K     & 5K       \\
2D Droplet (Fig.~\ref{fig:2D_droplet}(2nd row))   & 0.0108    & 1     & 16.4K     & 5K    \\
2D Droplet (Fig.~\ref{fig:2D_droplet}(3rd row))         & 0.0194    & 1     & 16.4K     & 5K       \\
2D Droplet (Fig.~\ref{fig:2D_droplet}(4th row))     & 0.0146    & 1     & 16.4K     & 5K       \\
2D Droplet (Fig.~\ref{fig:2D_droplet}(5th row))          & 0.0199    & 1     & 16.4K     & 5K       \\
Twisting bar (Fig.\ref{fig:bar_twisting}(1st row))        & 0.627     & 1     & 1.0M      & 193.3K   \\
Twisting bar (Fig.\ref{fig:bar_twisting}(2nd row))   & 0.624     & 1     & 1.0M      & 193.3K   \\
Twisting bar (Fig.\ref{fig:bar_twisting_inv_deformation}(1st row))        & 0.125     & 2  & 131.1K   & 48.3K \\
Twisting bar (Fig.\ref{fig:bar_twisting_inv_deformation}(2nd row))   & 0.124     & 2  & 131.1K   & 48.3K \\
Stretchy yo-yo (Fig.\ref{fig:strechy_bunny_1}(1st row))  & 1.314 &  2    &  4.2M     &  179.3K \\
Stretchy yo-yo (Fig.\ref{fig:strechy_bunny_1}(2nd row))     & 1.310 &  2    &  4.2M     &  179.3K \\
Snow bunny (Fig.\ref{fig:snow}) & 0.965     &  2    &  524.3K   &  116.3K \\
Snow bunny (EMPM in video)                 & 0.971     &  2    &  523.4K   &  116.3K \\
Water bunny (Fig.\ref{fig:water})       & 0.341     &  2    &  2.1M     &  96.3K \\
Water bunny (EMPM in video)                        & 0.381     &  2    &  2.1M     &  96.3K \\
FSI (Fig.\ref{fig:FSI}(1st row)) & 0.934         & 1 & 4.2M &  169.1K\\
FSI (Fig.\ref{fig:FSI}(2nd row))      & 0.939  & 1 & 4.2M & 169.1K\\
\hline
\end{tabular}
}
\end{center}
\label{tab:timings}
\vspace{-5mm}
\end{table}

\begin{table}[h!]
\caption{Configuration update cost in fluid-like simulations. \textbf{$\epsilon$} and  \textbf{$\eta$}: user-defined parameters to adjust the update frequency.  \textbf{$\tau$}: average update times per 104 time steps. \textbf{Cost}: average run-time.}
\vspace{-4mm}
\begin{center}
\setlength{\tabcolsep}{0.5mm}{
\begin{tabular}{lcccc}
\hline 
\textbf{Simulation} & \textbf{$\epsilon$} &  \textbf{$\eta$} &  \textbf{$\tau$} & \textbf{Cost} \\
\hline 
Snow bunny ($A$-ULMPM in Fig.\ref{fig:snow})          &   0.01      &0.01         & 24.33    &   0.112    \\
Snow bunny (EMPM in video)          & N/A        & N/A        &   104  &  0.479     \\
Water bunny ($A$-ULMPM in Fig.\ref{fig:water})          &  0.01       & 0.01        &  3.31   &     0.021  \\
Water bunny (EMPM in video)       & N/A        & N/A        &   104  &  0.662   \\
\hline
\end{tabular}
}
\end{center}
\label{tab:timings2}
\vspace{-5.5mm}
\end{table}

\subsection{2D Simulations for Solids and Fluids}
\subsubsection{Numerical fracture and cell-crossing instability}
We simulated a 2D rotating hyperelastic plate to demonstrate that EMPM suffers from the cell-crossing instability, which leads to severe numerical fractures, while TLMPM can completely eliminate these artifacts. Figure~\ref{fig:2D_rotating} shows that our $A$-ULMPM framework and TLMPM integrated with MLS-MPM captures the appealing hyperelastic rotation, preserving the angular momentum for long simulation periods while completely eliminating numerical fractures. However, traditional MLS-MPM~\cite{Hu:2018:Moving} that uses an Eulerian approach suffers from severe numerical fractures when large deformations occur. 

\begin{figure}[h!]
\vspace{-3mm}
\begin{overpic}[width=.18\textwidth]{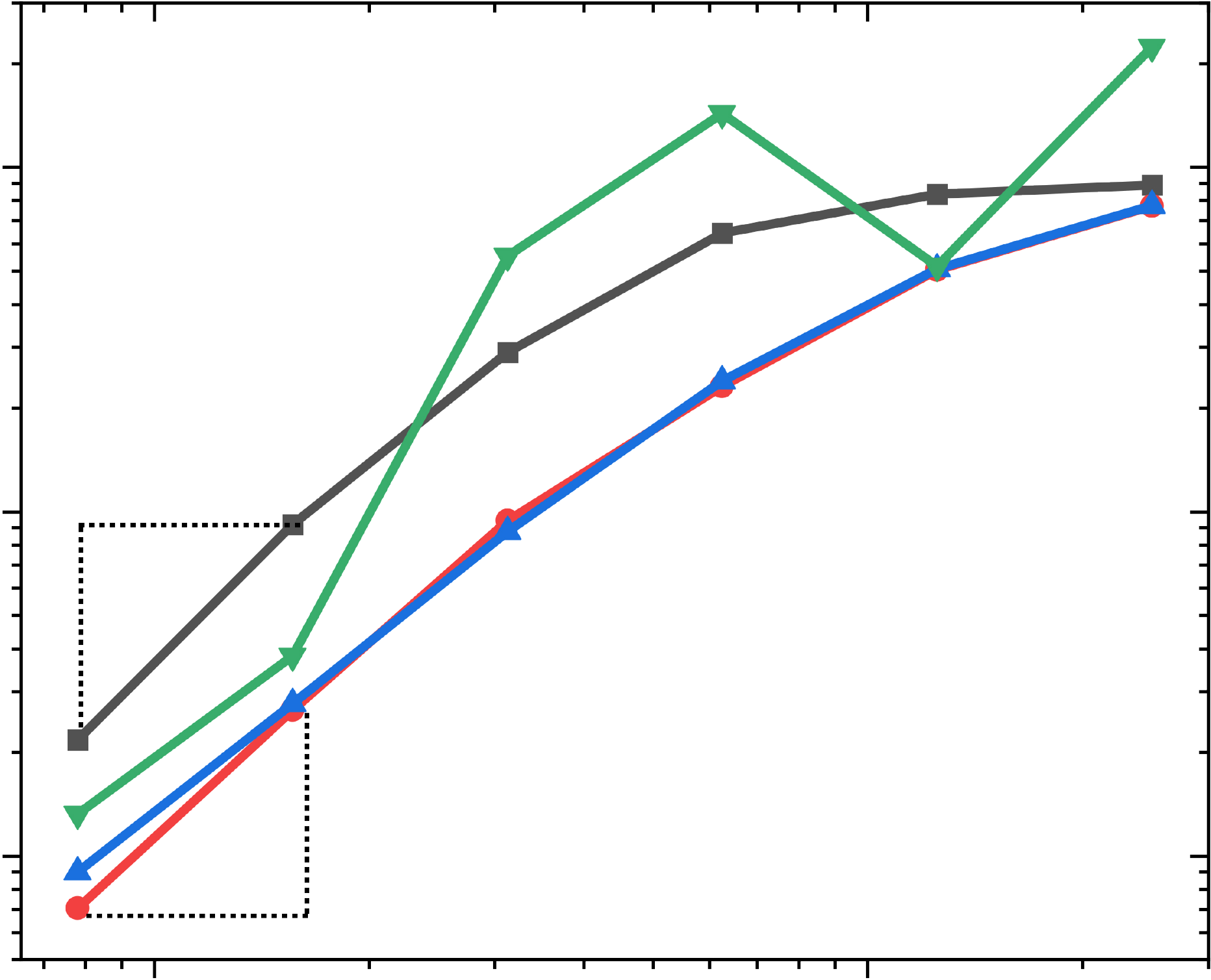}
    \put(30,10){\small slope$=2.08$}
    \put(40,-15){\small$\log_{2}(\Delta x)$}
    \put(-22,32){\small\rotatebox{90}{$\log_{2} (E)$}}
    \put(-19,0){\small $-17.6$}
    \put(-19,15){\small $-11.2$}
    \put(-19,30){\small $-10.2$}
    \put(-16,45){\small $-9.6$}
    \put(-16,60){\small $-9.2$}
    \put(-16,75){\small  $-8.7$}
    \put(0,-8){\small $-7$}
    \put(15,-8){\small $-6$}
    \put(35,-8){\small $-5$}
    \put(55,-8){\small $-4$}
    \put(75,-8){\small $-3$}
    \put(90,-8){\small $-2$}
\end{overpic}
\hspace{8mm}
\begin{overpic}[width=.18\textwidth]{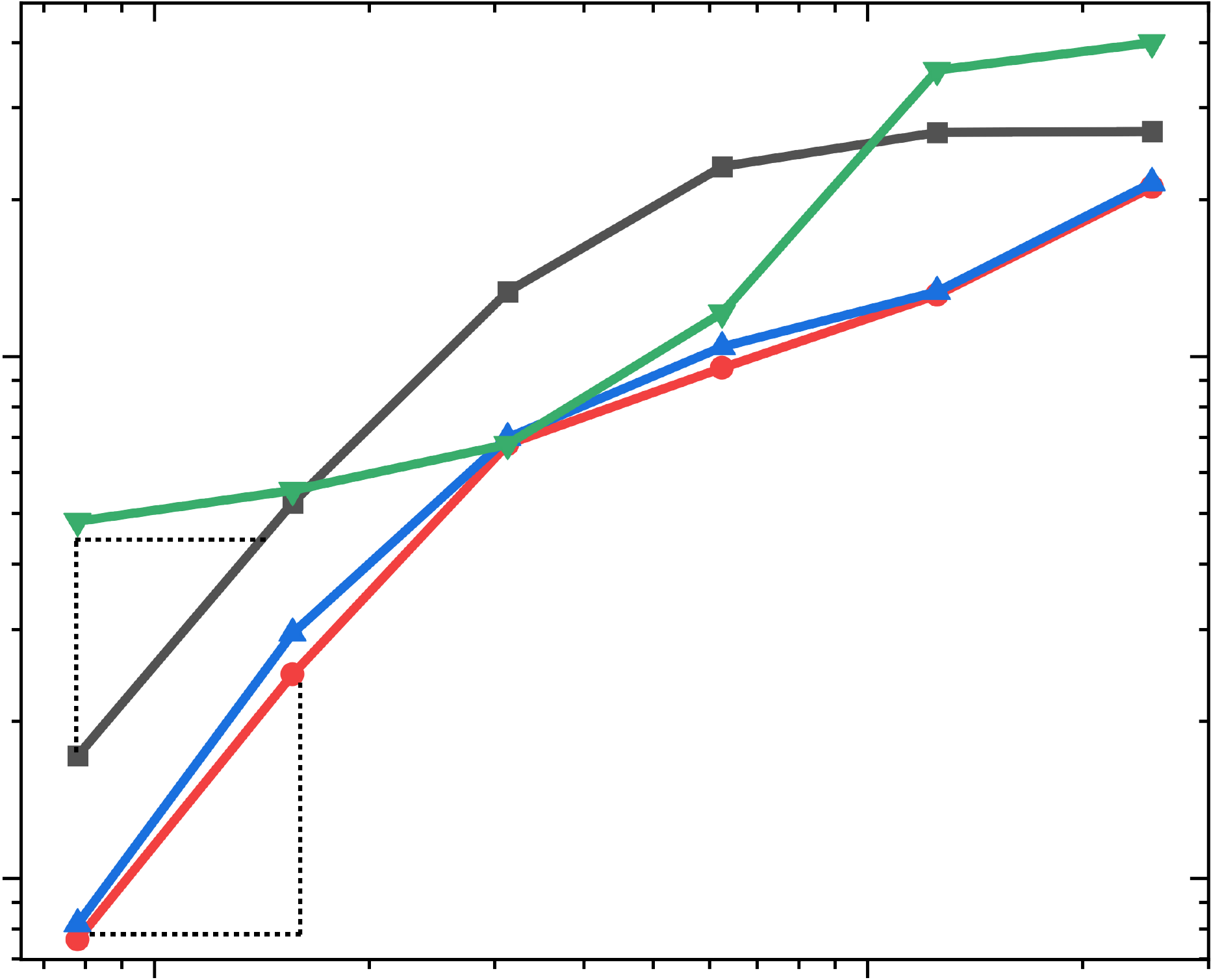} 
    \put(30,10){\small slope$=1.93$}
    \put(40,-15){\small$\log_{2}(\Delta x)$}
    \put(-22,30){\small\rotatebox{90}{$\log_{2} (E)$}}
    \put(-16,0){\small $-3.9$}
    \put(-16,15){\small $-2.9$}
    \put(-16,30){\small $-2.3$}
    \put(-16,45){\small $-1.9$}
    \put(-16,60){\small $-1.6$}
    \put(-16,75){\small  $-1.3$}
    \put(0,-8){\small $-7$}
    \put(15,-8){\small $-6$}
    \put(35,-8){\small $-5$}
    \put(55,-8){\small $-4$}
    \put(75,-8){\small $-3$}
    \put(90,-8){\small $-2$}
\end{overpic}
\vspace{3mm}
\caption{Log-log plots of error (labeled $E$) vs. grid mesh size $\Delta x$.
(Left) Error of the displacement, (right) error of the velocity using $A$-ULMPM integrated with MLS-MPM $(s=0)$ (red), $A$-ULMPM integrated with MLS-MPM $(s\neq 0)$ (blue), $A$-ULMPM integrated with MLS-MPM (black) $(s=n)$, and $A$-ULMPM integrated with kernel-MPM (black) $(s=0)$. Our $A$-ULMPM framework recovers TLMPM ($s=0$) and EMPM ($s=n$).   
Comparing the slopes and solution accuracy shows EMPM is non-convergent when grid-crossing occurs, while $A$-ULMPM with $s=0$ and $s\neq n$ provides convergent simulations with second order accuracy for displacement and velocity with small $\Delta x$.}
\label{fig:ConvPlot}
\vspace{-5mm}
\end{figure}
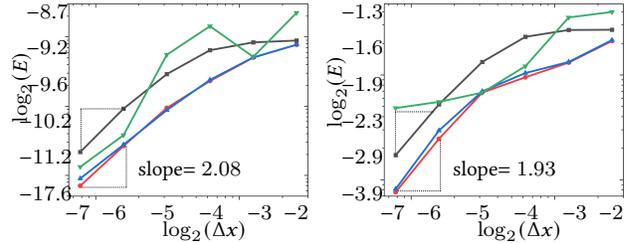

We quantitatively evaluate the spatial accuracy of our method on the rotating hyperelastic plate example by varying the grid resolution $\Delta x$. We utilized numerical results with resolution $256\times 256$ as the benchmark solution (see Figure~\ref{fig:2D_rotating}) and discretize sampling examples with spatial resolution $2^i\times2^i,\:i\in\{2,\ldots,7\}$ for the grid. 
We fixed the particle number $n_p=41943$ for each case and ran simulations up to a total simulation time of $0.025s$ with a time step size of $10^{-4}s$. The error for numerical simulations is defined as:
\begin{equation}
{E}=\sqrt{\frac{\left|\bs{\phi}_{i}-\bs{\phi}_{8}\right|^2}{n_p}}
    \label{eq:error}
\end{equation}
where $\bs{\phi}_{i}$ is the variable evaluated by lower grid resolutions, while $\bs{\phi}_8$ is the value at resolution $256^2$ (benchmark). 
Figure~\ref{fig:ConvPlot} shows the convergence plots for the average particle displacement and velocity for $A$-ULMPM integrated with MLS-MPM $(s=0)$, $A$-ULMPM integrated with MLS-MPM $(s\neq 0)$, $A$-ULMPM integrated with MLS-MPM (black) $(s=n)$, and $A$-ULMPM integrated with kernel-MPM $(s=0)$, where $A$-ULMPM ($s=0$) recovers TLMPM and $A$-ULMPM ($s=n$) recovers EMPM, as described in Section~\ref{sec:A_UL_MLS_MPM}. 
In general,  TLMPM can produce more accurate simulations since the cell-crossing error is completely eliminated, while EMPM fails to converge when the cell-crossing error is significant. $A$-ULMPM with $s\neq n $ exhibits second-order accuracy while its integration with MLS-MPM has higher solution accuracy than that with kernel-MPM.
\subsubsection{2D droplet}
We ran several simulations of falling droplets to compare our implementations in the $A$-ULMPM framework. As shown in Figure~\ref{fig:2D_droplet}, although TLMPM has benefits over EMPM for solid simulations (see Figure~\ref{fig:2D_rotating}), it fails to achieve detailed free surfaces in fluid simulations, as shown in Figure~\ref{fig:2D_droplet}, since the topology changes significantly in fluid-like simulations. 
$A$-ULMPM automatically updates configuration maps to produce similar fluid dynamics as EMPM and captures rich interactions. Moreover, our integration with MLS-MPM (see Section~\ref{sec:A_UL_MLS_MPM}) yields more energetic behavior compared to an integration with kernel-MPM (see Appendix~\ref{sec:A-ULMPM-kernel}). 

\subsection{Bar Twisting}
We imposed torsion and stretch boundary conditions at the two ends of a hyperelastic beam to showcase that $A$-ULMPM allows large deformations in solids without non-physical fractures. As shown in Figure~\ref{fig:bar_twisting}, traditional EMPM fails to preserve the shape of the beam during the twisting and pulling, while $A$-ULMPM can readily handle the challenging invertible elasticity when one end of the beam is released after twisting. Figure~\ref{fig:bar_twisting_inv_deformation} shows that $A$-ULMPM is capable of robustly recovering the beam shape after extreme elastic deformations, while particles in EMPM cluster into one irreversible (or plastic) thin string blocking the ``recovery''  of elasticity. 
\vspace{-1.5mm}
\subsection{Stretchy Yo-Yo} 
\begin{figure}[h!]
 \vspace{-4mm}
    \centering
    \includegraphics[width=\columnwidth]{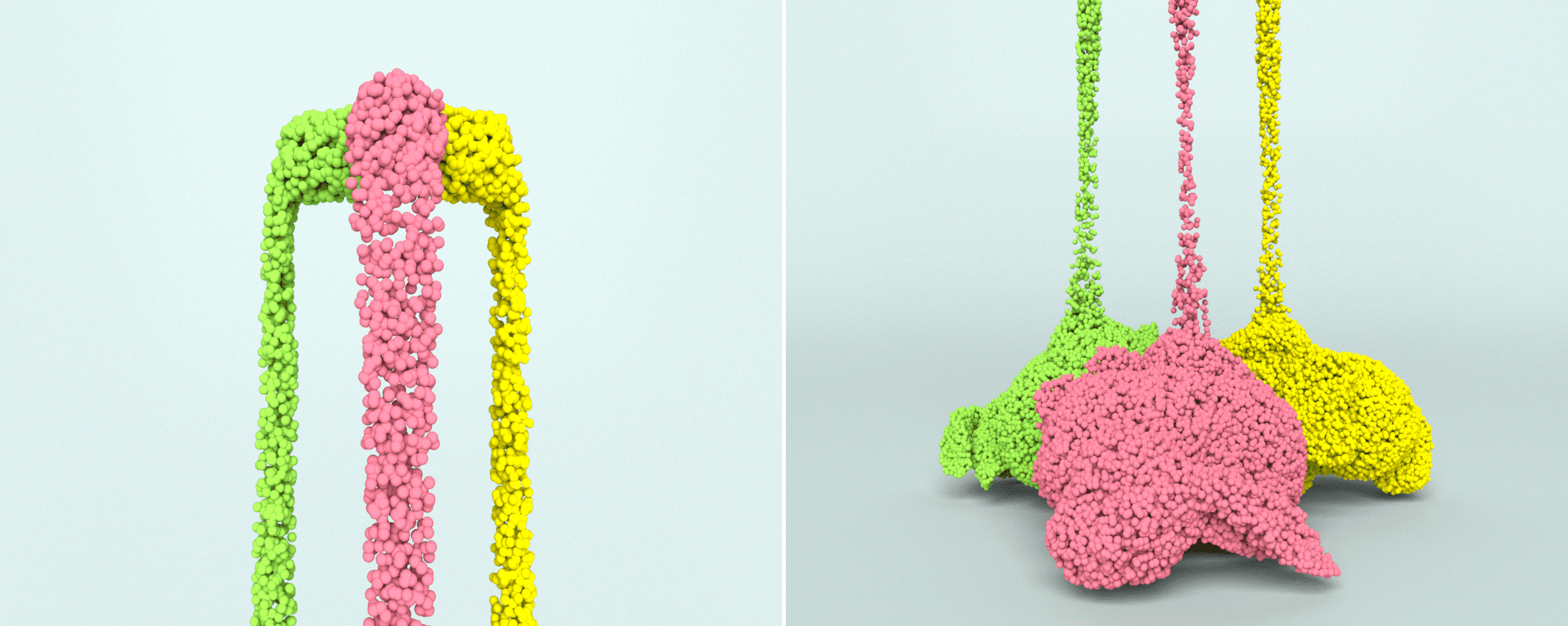}
     \vspace{-7.5mm}
    \caption{A closer view of the hyperelastic bunny yo-yo from Figure~\ref{fig:strechy_bunny_1}.}
    \label{fig:strechy_bunny_zoom_in}
    \vspace{-5mm}
\end{figure}

Next, we tossed a stretchy Stanford bunny yo-yo with gravity forces. Our $A$-ULMPM hyperelastic bunny demonstrates rich elastic responses and realistic bouncing dynamics, while EMPM breaks due to severe numerical fracture (see Figure~\ref{fig:strechy_bunny_1}). 
Our $A$-ULMPM framework can perfectly handle hyperelastic deformation under severe bending (see Figure~\ref{fig:strechy_bunny_zoom_in} (left)) and stretching (see Figure~\ref{fig:strechy_bunny_zoom_in} (right)).
% \vspace{-1.5mm}
\subsection{Fluid-like Bunny Simulation} 
We simulated the fluid-like behavior of different materials using our $A$-ULMPM framework, as described in Section~\ref{sec:A_UL_MLS_MPM}, such as elastoplastic snow~\cite{Stomakhin:2013:MPMsnow} and weakly compressible water. 
We dropped two copies of the snow Stanford bunny with different orientations to a solid wedge, as shown in Figure~\ref{fig:snow}. $A$-ULMPM captures the vivid snow smashing and scattering after the bunnies fall on the wedge, similar to its Eulerian counterparts proposed in prior works~\cite{Stomakhin:2013:MPMsnow} (see side-by-side comparisons in our video). 
We simulated a water Stanford bunny falling inside a spherical container (see Figure~\ref{fig:water}), showing the extreme large deformations and energetic splashes.   
 $A$-ULMPM does not update the configuration maps at each time step, in contrast to EMPM, so it is naturally more efficient in fluid-like simulations (see Table~\ref{tab:timings2}).
 
\subsection{Fluid-Structure Interaction}

Our $A$-ULMPM framework can also simulate realistic fluid-structure interaction problems, as shown in Figure~\ref{fig:FSI}.
\begin{figure}[h!]
\vspace{-4mm}
    \centering
    \includegraphics[width=.49\columnwidth]{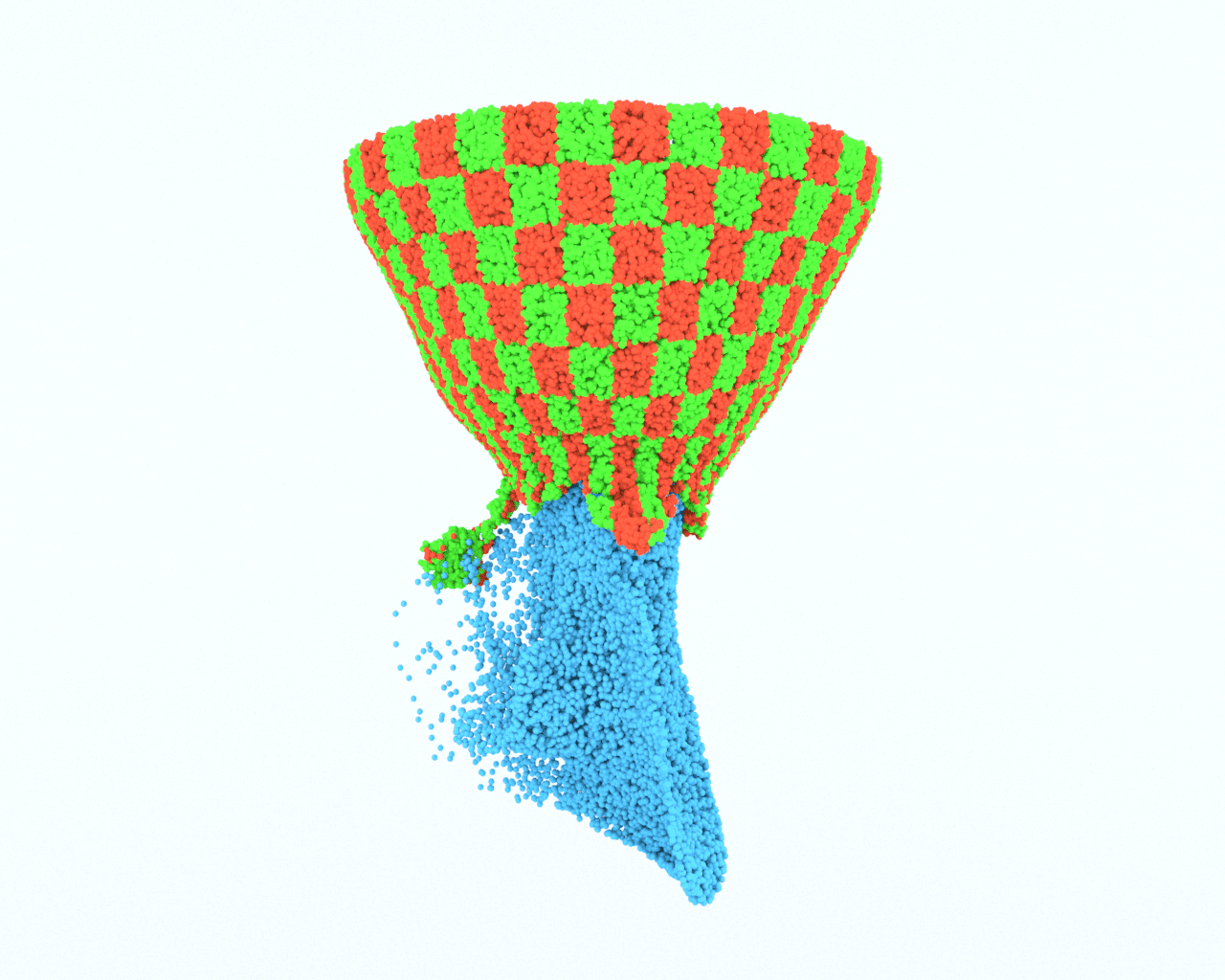}
    \includegraphics[width=.49\columnwidth]{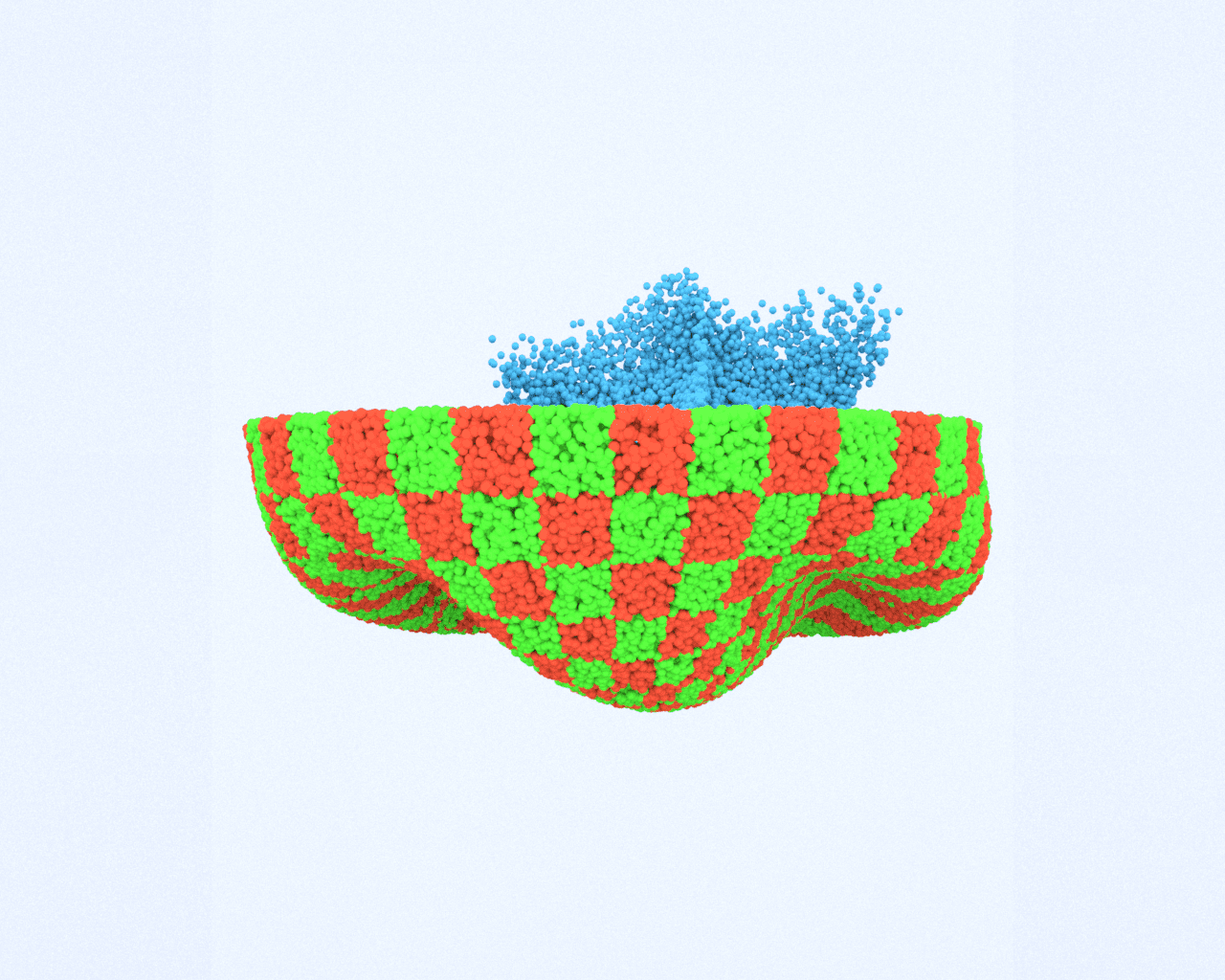}
    \vspace{-3mm}
    \caption{A closer view of hyperelastic deformations in the fluid-structure interaction example from Figure~\ref{fig:FSI}. (Left) The bowl breaks in EMPM due to numerical fractures, while (right) $A$-ULMPM maintains a leakproof bowl. }
        \vspace{-5mm}
    \label{fig:fsi_zoom_in}
\end{figure}
\vspace{1.5mm}
We dropped a water Stanford bunny to a hyperelastic bowl. $A$-ULMPM captures rich fluid-structure interactions, displaying advantages of both TLMPM and EMPM (see Figure~\ref{fig:fsi_zoom_in} (right)), and highlighting that $A$-ULMPM can adaptively handle both fluid and solid simulations without having to switch between TLMPM and EMPM.
In contrast, EMPM suffers from numerical fractures. As shown in Figure~\ref{fig:FSI}, the hyperelastic bowl fractures, causing the water to spill below.

\section{Discussion and Conclusion}
\label{sec:conclusion}

\subsection{Limitations and Future Work} 
Our model has generated a large number of compelling examples, but there remains much work to be done. Parameters to adjust the configuration update frequency were tuned by hand, and it would be interesting to calibrate them to measured models. Since each object has it own configuration map, we only briefly investigated contact by projecting particle positions to a global grid and processing grid-based collisions~\cite{Stomakhin:2013:MPMsnow}. By doing this, we observed slight self-penetration in the 3D twisting beam example (see Figure~\ref{fig:bar_twisting}). This could be addressed by further introducing contact algorithms, such as~\cite{Jiang:2017:Anisotropic, Han:2019:Hybrid}.
While we did not experience a need for excessively small time steps given the low stiffness of the materials we considered, deforming materials with high wave speed, such as steel, could benefit from a fully implicit discretization.
Though side-to-side comparisons with Eulerian MLS-MPM shows that our $A$-ULMPM framework produces similar fluid-like dynamics as EMPM and captures rich interactions, a slight energy loss in $A$-ULMPM framework was observed. 
This could be addressed by introducing more accurate configuration update criteria. 
Finally, while our focus was on the material responses of water, snow, and hyperelastic solids, it would be interesting to investigate other materials and multi-physics coupling problems with $A$-ULMPM. 

\subsection{Conclusion}
We proposed an arbitrary updated Lagrangian Material Point Method ($A$-ULMPM), which combines advantages of Total Lagrangian frameworks~\cite{De:2021:TLMPM,De:2020:TLMPM} and Eulerian frameworks~\cite{Stomakhin:2013:MPMsnow, Hu:2018:Moving} in an adaptive fashion. $A$-ULMPM avoids the cell-crossing instability and numerical fracture in solid simulations, while still allowing for large deformations that arise in fluid-like simulations. 
It can be easily integrated with any existing MPM framework and builds a foundation for devising various MPM schemes, such as PolyPIC~\cite{Fu:2017:PolyPIC}, for enhanced accuracy and visual vividness.

%Future work can explore many interesting avenues. Parameters to adjust the configuration update frequency were tuned by hand, and it would be interesting to calibrate them to measured models. Since each object has it own configuration map, we only briefly investigated contact by projecting particle positions to a global grid and processing grid-based collisions~\cite{Stomakhin:2013:MPMsnow}. By doing this, we observed slight self-penetration in the 3D twisting beam example (see Figure~\ref{fig:bar_twisting}). This could be addressed by further introducing contact algorithms, such as~\cite{Jiang:2017:Anisotropic, Han:2019:Hybrid}.
%While we did not experience a need for excessively small time steps given the low stiffness of the materials we considered, deforming materials with high wave speed, such as steel, could benefit from a fully implicit discretization. 
%Finally, while our focus was on the material responses of water, snow, and hyperelastic solids, it would be interesting to investigate other materials and multi-physics coupling problems with $A$-ULMPM. 
\bibliographystyle{ACM-Reference-Format}
\bibliography{Reference}
\appendix
\label{sec:appendix}
\section{MLS Gradient}
\label{sec:mls_gradient}
Consider a domain {($\Omega_0$)} in Euclidean space $\mathbb{E}$. Let $\bs{q}_i\in \Omega_{0} \subset \mathbb{E}$ denote the geometric position of a certain point in the domain and $\bs{q}_j\in \Omega_{0} \subset \mathbb{E}$ denote another point which is close to the position $\bs{q}_i$.
Therefore, any physical quantity at each point in space can be defined as $\bs{\varphi}(\bs{q}_i)$ and following a first-order Taylor approximation gives: 
\begin{equation}
\bs{\varphi}(\bs{q}_j)\cong \bs{\varphi}(\bs{q}_i) +\nabla{\bs{\varphi}}_{ij} \bs{r}_{ij}
\label{eq:Taylor1st}
\end{equation}
where $\bs{r}_{ij}=\bs{q}_j-\bs{q}_i$. A weighted error $E_i$ of equation~\eqref{eq:Taylor1st} is defined within its associated domain $\Omega_i$ as follows:
\begin{equation}
E_i=\int_{j\in\Omega_i}\|{\bs{\varphi}(\bs{q}_j)-\bs{\varphi}(\bs{q}_i)}-\nabla{\bs{\varphi}_{ij}} \bs{r}_{ij}\|^2W_{ij}dV
\label{eq:error_i}
\end{equation}
where $\Omega_i$ represents the neighboring domain of $\bs{q}_i$; $W_{ij}$ represents the influence from $j$ to $i$ and is a function of $\left\|\bs{r}_{ij}\right\|$. To minimize the error $E_i$ and get the best approximation, the Least Squares Technique is exploited by introducing $\frac{\partial E_i}{\partial \nabla \bs{\phi}_i}=0$:
\begin{equation}
\frac{\partial}{\partial\nabla \bs{\phi}_i}\int_{j\in\Omega_i}\left({\bs{\varphi}(\bs{q}_j)-\bs{\varphi}(\bs{q}_i)}-\nabla{\bs{\varphi}_{ij}} \bs{r}_{ij}\right)^2W_{ij}dV=0 
\end{equation}
Therefore, we have a MLS-gradient operator : 
\begin{eqnarray}
\nabla\bs{\varphi}(\bs{q}_i)=\left(\int_{j\in\Omega_i}\bs{\varphi}(\bs{q})_{ij}\otimes \bs{r}_{ij} W_{ij} dV\right)\left(\int_{j\in\Omega_i}\bs{r}_{ij}\otimes\bs{r}_{ij} W_{ij} dV \right)^{-1}
\label{eq:gradient_int}
\end{eqnarray}
where $\bs{\varphi}(\bs{q})_{ij}=\bs{\varphi}(\bs{q})_{j}-\bs{\varphi}(\bs{q})_{i}$
and its discrete form is given by:
\begin{eqnarray}
\nabla\bs{\varphi}(\bs{q}_i)=\left(\sum_{j\in\Omega_i}\bs{\varphi}(\bs{q})_{ij}\otimes \bs{r}_{ij} W_{ij} V_j\right)\left(\sum_{j\in\Omega_i}\bs{r}_{ij}\otimes\bs{r}_{ij} W_{ij} V_j \right)^{-1}
\label{eq:gradient_sum}
\end{eqnarray}
where $V_j$ denotes volume distributed at $\bs{q}_j$. 
Moreover, the derivation of $\nabla\bs{\varphi}(\bs{q}_i)$ with respect to $\bs{q}_k$ is given as follows:
\begin{eqnarray}
\begin{split}
&\frac{\p\nabla\bs{\varphi}(\bs{q}_i)}{\p \bs{\varphi}(\bs{q}_k)}=\left(\sum_{j\in\Omega_i}\frac{\bs{\varphi}(\bs{q})_{ij}}{\p\bs{\varphi}(\bs{q}_{k})}\otimes \bs{r}_{ij} W_{ij} V_j\right)\left(\sum_{j\in\Omega_i}\bs{r}_{ij}\otimes\bs{r}_{ij} W_{ij} V_j \right)^{-1} \\
&=\left(\sum_{j\in\Omega_i}\left(\delta_{jk}-\delta_{ik}\right)\otimes \bs{r}_{ij} W_{ij} V_j\right)\left(\sum_{j\in\Omega_i}\bs{r}_{ij}\otimes\bs{r}_{ij} W_{ij} V_j \right)^{-1}\\
\end{split}
\label{eq:dFdq}
\end{eqnarray}
where $\delta$ is the Kronecker delta function.

\section{$A$-ULMPM Integration with kernel-MPM}
\label{sec:A-ULMPM-kernel}
We briefly describe the integration of our $A$-ULMPM with traditional MPM~\cite{Stomakhin:2013:MPMsnow} as follows:
\begin{enumerate}
\item{Transfer Particle to Grid.}
\begin{equation}
    m_i^s \bs{v}_i^n=\sum_p m_p \bs{v}_p^n W_{pi}^s,\quad m_i^s=\sum_p m_p W_{pi}^s,\quad \bs{v}_i^n= \frac{m_i^s \bs{v}_i^n}{m_i}
\end{equation}
\item{Update Grid Momentum.} 
\begin{equation}
\begin{split}
    &\bs{\hat{v}}_i^{n+1}=\bs{v}_i^n-\frac{\dt}{m_i}\sum_p\mathbb{P}_p^0(\bs{F}_p^{0s})^T \nabla^sW_{ip}^s V_j^0\\
    &\bs{q}_i^{n+1}=\bs{q}_i^n+\dt\bs{\hat{v}}_i^{n+1}\\
\end{split}
\label{eq:kernel_grid_Vel_update_explicit}
\end{equation}
\item{Transfer Grid Velocity to Particles.} 
We project $\bs{q}_i^{n+1}$ for all objects to the global grid and consider grid-based collisions~\cite{Stomakhin:2013:MPMsnow} to obtain $\bs{v}_i^{n+1}$. We transfer $\bs{v}_i^{n+1}$ to particles using a weighted combination of PIC and FLIP, as described in~\cite{Stomakhin:2013:MPMsnow}. 
\item{Update Particle Deformation Gradient.}
\begin{eqnarray}
  \bs{F}_p^{s(n+1)}=    &\bs{F}_p^{sn}+\dt\nabla_s\bs{v}_p^{n+1}\\
\label{eq:kernel_gradient_s(n+1)}
\end{eqnarray}
where $\nabla_s\bs{v}_p^{n+1}$ is given by:

\begin{equation}
\nabla_s\bs{v}_p^{n+1}=\sum_i\bs{v}_i^{n+1} \nabla^sW_{pi}^s
\label{eq:kernelgradientVp}
\end{equation}
By differential chain rule, the update of $\bs{F}_p^{0(n+1)}$ is given by:
\begin{equation}
    \bs{F}_p^{0(n+1)}=\frac{\p \bs{q}_p^{n+1}}{\p\bs{q}_p^{s}}\frac{\p\bs{q}_p^{s}}{\p\bs{q}_p^{0}}
    =\bs{F}_p^{s(n+1)}\bs{F}_p^{0s}
\end{equation}
\end{enumerate}

\end{document}